\documentclass[twocolumn]{aastex631}

\shortauthors{Pedrini et al.}
\graphicspath{{./}{figures/}}

\begin{document}

\title{Feedback in Emerging extragAlactic Star clusTers, FEAST: JWST spots PAH destruction in NGC 628 during the emerging phase of star formation}

\correspondingauthor{Alex Pedrini}
\email{alex.pedrini@astro.su.se}

\author[0000-0002-8222-8986]{Alex Pedrini}
\affiliation{Department of Astronomy, Oskar Klein center, Stockholm University, AlbaNova University center, SE-106 91 Stockholm, Sweden}

\author[0000-0002-8192-8091]{Angela Adamo}
\affiliation{Department of Astronomy, Oskar Klein center, Stockholm University, AlbaNova University center, SE-106 91 Stockholm, Sweden}

\author[0000-0002-5189-8004]{Daniela Calzetti}
\affiliation{Department of Astronomy, University of Massachusetts, 710 North Pleasant Street, Amherst, MA 01003, USA}

\author[0000-0001-8068-0891]{Arjan Bik}
\affiliation{Department of Astronomy, Oskar Klein center, Stockholm University, AlbaNova University center, SE-106 91 Stockholm, Sweden}

\author[0000-0003-4910-8939]{Benjamin Gregg}
\affiliation{Department of Astronomy, University of Massachusetts, 710 North Pleasant Street, Amherst, MA 01003, USA}

\author[0000-0002-1000-6081]{Sean T. Linden}
\affiliation{Department of Astronomy, University of Massachusetts, 710 North Pleasant Street, Amherst, MA 01003, USA}
\affiliation{Steward Observatory, University of Arizona, 933 N. Cherry Avenue, Tucson, AZ 85719, USA}

\author[0009-0008-4009-3391]{Varun Bajaj}
\affiliation{Space Telescope Science Institute, 3700 San Martin Drive Baltimore, MD 21218, USA}

\author[0000-0002-2918-7417]{Jenna E. Ryon}
\affiliation{Space Telescope Science Institute, 3700 San Martin Drive Baltimore, MD 21218, USA}

\author[0000-0001-5189-4022]{Ahmad A. Ali}
\affiliation{I. Physikalisches Institut, Universität zu Köln, Zülpicher Str. 77, 50937 Köln, Germany}

\author[0009-0003-6182-8928]{Giacomo Bortolini}
\affiliation{Department of Astronomy, Oskar Klein center, Stockholm University, AlbaNova University center, SE-106 91 Stockholm, Sweden}

\author[0000-0001-6464-3257]{Matteo Correnti}
\affiliation{INAF Osservatorio Astronomico di Roma, Via Frascati 33, 00078, Monteporzio Catone, Rome, Italy}
\affiliation{ASI-Space Science Data Center, Via del Politecnico, I-00133, Rome, Italy}

\author[0000-0002-1723-6330]{Bruce G. Elmegreen}
\affiliation{Katonah, NY 10536}

\author[0000-0002-1392-3520]{Debra Meloy Elmegreen}
\affiliation{Vassar College, Department of Physics and Astronomy, Poughkeepsie, NY 12604, USA}

\author[0000-0001-8608-0408]{John S. Gallagher}
\affiliation{Department of Astronomy, University of Wisconsin-Madison, 475 N. Charter Street, Madison, WI 53706, USA}

\author[0000-0002-3247-5321]{Kathryn~Grasha}
\altaffiliation{ARC DECRA Fellow}
\affiliation{Research School of Astronomy and Astrophysics, Australian National University, Canberra, ACT 2611, Australia} 
\affiliation{ARC center of Excellence for All Sky Astrophysics in 3 Dimensions (ASTRO 3D), Australia}

\author[0000-0002-6447-899X]{Robert A. Gutermuth}
\affiliation{Department of Astronomy, University of Massachusetts, 710 North Pleasant Street, Amherst, MA 01003, USA}

\author[0000-0001-8348-2671]{Kelsey E. Johnson}
\affiliation{Department of Astronomy, University of Virginia, Charlottesville, VA 22904, USA}

\author[0000-0003-0470-8754]{Jens Melinder}
\affiliation{Department of Astronomy, Oskar Klein center, Stockholm University, AlbaNova University center, SE-106 91 Stockholm, Sweden}

\author[0000-0003-1427-2456]{Matteo Messa}
\affiliation{INAF - Osservatorio di Astrofisica e Scienza dello Spazio di Bologna, Via Gobetti 93/3, I-40129 Bologna, Italy}

\author[0000-0002-3005-1349]{Göran Östlin}
\affiliation{Department of Astronomy, Oskar Klein center, Stockholm University, AlbaNova University center, SE-106 91 Stockholm, Sweden}

\author[0000-0003-2954-7643]{Elena Sabbi}
\affiliation{Space Telescope Science Institute, 3700 San Martin Drive Baltimore, MD 21218, USA}
\affiliation{Gemini Observatory/NSF’s NOIRLab, 950 North Cherry Avenue, Tucson, AZ, 85719, USA}

\author[0000-0002-0806-168X]{Linda J. Smith}
\affiliation{Space Telescope Science Institute, 3700 San Martin Drive Baltimore, MD 21218, USA}

\author[0000-0002-0986-4759]{Monica Tosi}
\affiliation{INAF - Osservatorio di Astrofisica e Scienza dello Spazio di Bologna, Via Gobetti 93/3, I-40129 Bologna, Italy}

\author[0000-0002-2199-0977]{Helena Faustino Vieira}
\affiliation{Cardiff Hub for Astrophysics Research and Technology (CHART), School of Physics \& Astronomy, Cardiff University, The Parade,
Cardiff CF24 3AA, UK}









\begin{abstract}
We investigate the emergence phase of young star clusters in the nearby spiral galaxy NGC 628. We use JWST NIRCam and MIRI observations to create spatially resolved maps of the Pa$\alpha$-1.87 $\mu$m and Br$\alpha$-4.05 $\mu$m hydrogen recombination lines, as well as the 3.3 $\mu$m and 7.7 $\mu$m emission from polycyclic aromatic hydrocarbons (PAHs). We extract  953 compact HII regions and analyze the PAH emission and morphology at $\sim$10 pc scales in the associated photo-dissociation regions (PDRs). While HII regions remain compact, radial profiles help us to define three PAH morphological classes: compact ($\sim$ 42 \%), extended ($\sim$ 34 \%) and open ($\sim$ 24 \%). The majority of compact and extended PAH morphologies are associated with very young star clusters ($<$5 Myr), while open PAH morphologies are mainly associated with star clusters older than 3 Myr. 
We observe a general decrease in the 3.3 $\mu$m and 7.7 $\mu$m PAH band emission as a function of cluster age, while their ratio remains constant with age out to 10 Myr and morphological class. The recovered PAH$_{3.3 \mu{\rm m}}$/PAH$_{7.7 \mu{\rm m}}$ ratio is lower than values reported in the literature for reference models that consider neutral and ionized PAH populations and analyses conducted at galactic scales. The 3.3 $\mu$m and 7.7 $\mu$m bands are typically associated to neutral and ionised PAHs, respectively. While we expected neutral PAHs to be suppressed in proximity of the ionizing source, the constant PAH$_{3.3 \mu{\rm m}}$/PAH$_{7.7 \mu{\rm m}}$ ratio would indicate that both families of molecules disrupt at similar rates in proximity of the HII regions.


\end{abstract}



\section{Introduction} \label{sec:intro}
Star formation within galaxies operates across a wide range of physical scales, from stellar aggregates at sub-parsec sizes to kilo-parsec scales encompassing the interstellar medium (ISM)  \citep[e.g.,][]{Efremov95, Elmegreen06}. At parsec scales, we find young star clusters that form in the densest cores of giant molecular clouds (GMCs). Emerging young star clusters (eYSCs) have been broadly studied in the Milky Way especially at MIR and radio wavelengths \citep[e.g.,][]{Camargo15,Ramirez16,Stutz18} and, in a few cases, in nearby galaxies by means of HST UV-to-NIR and radio observations \citep[e.g.,][in IC 4662, NGC 1705, NGC 5398, NGC 1313 and the Antennae galaxy, respectively]{Johnson03, Johnson15, Messa21, he2022}. The emergence of young star clusters paves the way for the complex and rich stellar distributions that we observe in galaxies \citep{Kennicutt12}. Very energetic photons ($>$ 13.6 eV) produced in young and massive OB stars in and around eYSCs ionize the surrounding gas, powering an ionization front at the edge of the HII region, which rapidly expands \citep{Osterbrock06, Deharveng10, Draine11}. Outside the ionization front, less energetic photons (5 - 13.6 eV) interact with the neutral gas and molecules, producing photo-dissociation regions \citep[PDRs,][]{Hollenbach99}. 

Within the PDRs, dust grains absorb and reprocess stellar photons and act as an important cooling source for the ISM. 
In the very small grains regime, we find the Polycyclic Aromatic Hydrocarbons (PAH). In this family of hydrocarbons, carbon atoms are organized in planar hexagonal rings and hydrogen atoms lie at the boundary of the rings \citep[][and references therein]{Tielens08}.
PAH molecules absorb UV photons, triggering vibrational de-excitation through IR emission \citep{Leger89}.
Their emission dominates the global IR spectrum of star-forming galaxies, representing up to $\sim$ 20\% of their total IR luminosity \citep[e.g.][]{Smith07, Tielens08, Lai20}.

PAH spectral features arise at 3.3, 6.2, 7.7, 8.6, 11.3, 12.7, 16.4, and 17 $\mu$m and are linked to the different vibrational modes and composition of their molecules \citep{Draine21,Chown23}. Their emission properties strongly depend on their sizes (e.g. the number of carbon atoms), the internal temperature \citep{Leger89}, and the different ionization states of the PAH molecules \citep{Draine07, Draine21}. Recently, \cite{Peeters23} presented new JWST observations of PAH emission in the Orion Bar. This work confirmed that PAH emission is an excellent tracer for the atomic PDR. Moreover, they provided new important insights into the photo-chemical evolution of PAHs.  
\\
The excitation of PAHs is primarily driven by UV and optical photons emitted from stars. As a result, young star-forming regions provide higher heating capability and therefore PAHs are brighter in regions where star formation is actively ongoing. PAH emission is thus co-located with star formation, making these molecules valuable as tracers for the star formation rate in galaxies (see summary in \citealt{Calzetti13}).
On galactic scales, several studies in the literature \citep[e.g.][]{Xie19,Mallory22} have estimated global star formation rates in local galaxies using PAH tracers. 

However, at GMCs scales, \cite{Povich07} (among many others) observed a decrease of PAH emission inside star-forming regions in the Milky Way at physical scales of $\sim 5$ pc. \cite{Chastenet23B} obtained similar results looking at nearby galaxies, measuring the PAH fraction at the scales of star-forming complexes (50-100 pc). 

There are several mechanisms that could be responsible for the observed lack of PAH emission in star-forming regions \citep[][and references therein]{Li20}.
Interstellar shocks and coagulation onto dust grains might lead to the loss of an important PAH fraction of 20-40\% in carbon atoms \citep{Seok14}. Moreover, \cite{Micelotta10} showed that shocks can also destroy the grains. These mechanisms play a crucial role in the evolution of PAHs and they are strictly connected to the physical processes that lead to the formation of PAHs in star-forming regions. 

Another commonly accepted scenario is that energetic UV photons can destroy PAHs \citep[photo-termo-dissociation][]{Leger89b} within a star-forming region \citep[e.g.,][]{Guhathakurta89,Allain96,Verstraete01,Povich07,Watson08,Churchwell09,Xie22,Egorov23}. 
\cite{Engelbracht05,Lebouteiller11} and \cite{Sandstrom12} observed a decrease of PAH emission in low metallicity environments, which could be related to the hardness of the radiation fields \citep{Gordon08}.
Furthermore, \cite{Egorov23} found a strong anti-correlation between the quantity R$_{\rm PAH}$, defined as R$_{\rm PAH} =$ (F770W + F1130W)/F2100W, and the ionization parameter in HII regions of nearby galaxies, confirming that the survival of PAH molecules is related to the properties of the radiation field and the evolutionary stage of the star-forming regions.

Despite the many steps taken towards a better understanding of this mechanism, we have not yet discovered the role played by PAH destruction in the emerging phase of star formation, when eYSCs have not yet cleared the natal cloud \citep[e.g.,][]{Krumholz19}. Milky Way observations of star-forming regions can reach high resolutions, but superposition effects make the interpretation of the results challenging. Conversely, observations of PAHs in star-forming regions outside the Local Volume before the advent of JWST never achieved the resolution needed to study the morphology of these regions and compare them with the emission from the HII regions.
During the last two decades, H$\alpha$-HST observations have been extensively used to study the morphology of HII regions during the young-intermediate optically-visible phases of star formation, defining an evolution based on the morphological variation of the ionized gas \citep{Whitmore11,Hollyhead15,Hannon19,Hannon22,Messa21,Calzetti23}. 
With JWST, we can now trace the spatial scales needed to study the very young emerging phase of star formation, where HII regions are compact and the stellar feedback from massive stars is actively shaping the PDRs and the PAHs that reside in them. 
\\

In this paper, we present a study of star-forming regions at $\sim 7-10$ parsecs scale in the spiral galaxy NGC 628 (M74). NGC 628 is a main-sequence star-forming galaxy at a distance of 9.84 Mpc \citep{Tully09}. The absence of intense nuclear star formation and the presence of an intricate complex of dusty filaments along the spiral arms make this galaxy a unique laboratory to study the morphology of the star-forming regions \citep{Elmegreen19}. 

Recently, the PHANGS collaboration (PHANGS-JWST program, \citealt{Lee23}) presented new results on the PAH emission in star-forming regions at $\sim$ 50 pc scales in NGC 628. \cite{Watkins23} identified stellar feedback-driven bubbles in the 7.7 $\mu$m PAH emission feature and statistically quantified them. This analysis evidenced a complex distribution of bubbles, suggesting that bubble merging is a common process and that it is important in setting constraints on the global properties of the galaxy. For Milky Way bubbles, this has been quantified thanks to the Spitzer's GLIMPSE survey \citep[e.g.][]{Churchwell06}.
Furthermore, \cite{Chastenet23B} and \cite{Sutter24} studied the coincidence of PAH emission and ionized gas at $\sim 1\arcsec$ resolution, witnessing a general decrease of R$_{\rm PAH}$ at the location of the HII regions \citep{Emsellem22}.

In this study, we further zoom into star-forming regions, reaching scales of a few parsecs. We can now unveil the evolution of PDRs during the first stages of star formation, where HII regions are still very bright and compact. 
To achieve this goal, we take advantage of the exquisite resolution of JWST using NIRCam and MIRI observations of NGC 628 obtained by the Feedback in Emerging extragAlactic Star clusTers (FEAST) program (GO 1783, PI A.Adamo). 
Particularly, we used the 3.3 $\mu$m emission band to infer the properties of the PAHs. This emission feature traces the small and neutral PAHs that arise from a C-H stretching vibration \citep{Maragkoudakis20}.

The manuscript is organized as follows. After a presentation of the observations of NGC 628 and the adopted data reduction procedures in Sec.~\ref{sec:obs&dr}, we describe the methodology utilized to characterize HII regions and PDRs as traced by the 3.3 $\mu$m PAH emission and we present the PAH-based morphological classification of star-forming regions in Sec.~\ref{sec:methods}.
We present the results of our analysis in Sec.~\ref{sec:res} and we discuss them in Sec.~\ref{sec:discussion}. Finally, we summarize our findings in Sec.~\ref{sec:end}. 


\section{Observations and data reduction} \label{sec:obs&dr}

\begin{figure*}[htb]
    \centering
    \includegraphics[width = \textwidth]{RGB_4paper_checkpoint_afterFlatten_ROTATED_CROPPED_crops_scale_arrow2_compressed.pdf}
    \\[\smallskipamount]
    \includegraphics[width = 0.495\textwidth]{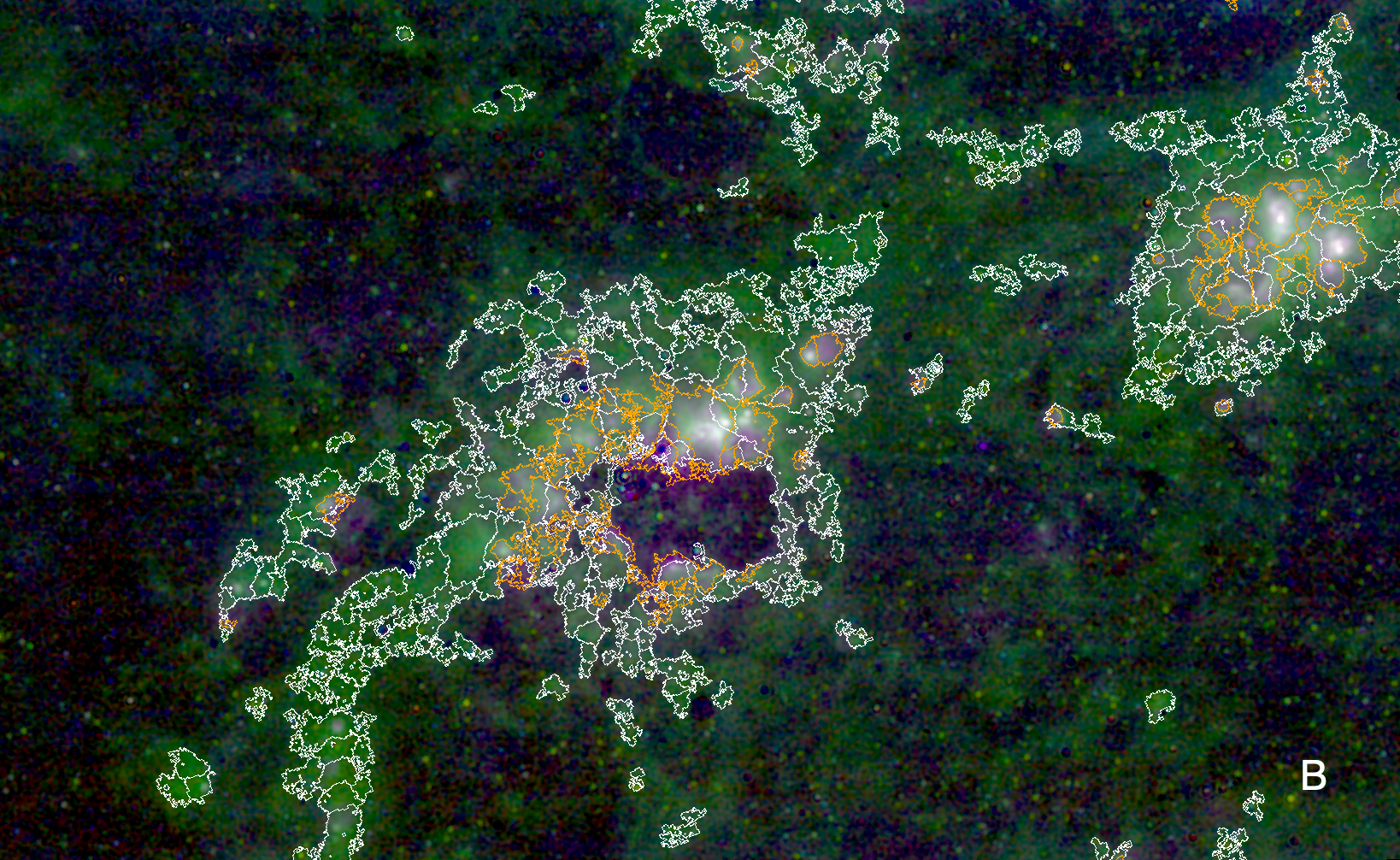}\hfill
    \includegraphics[width = 0.495\textwidth]{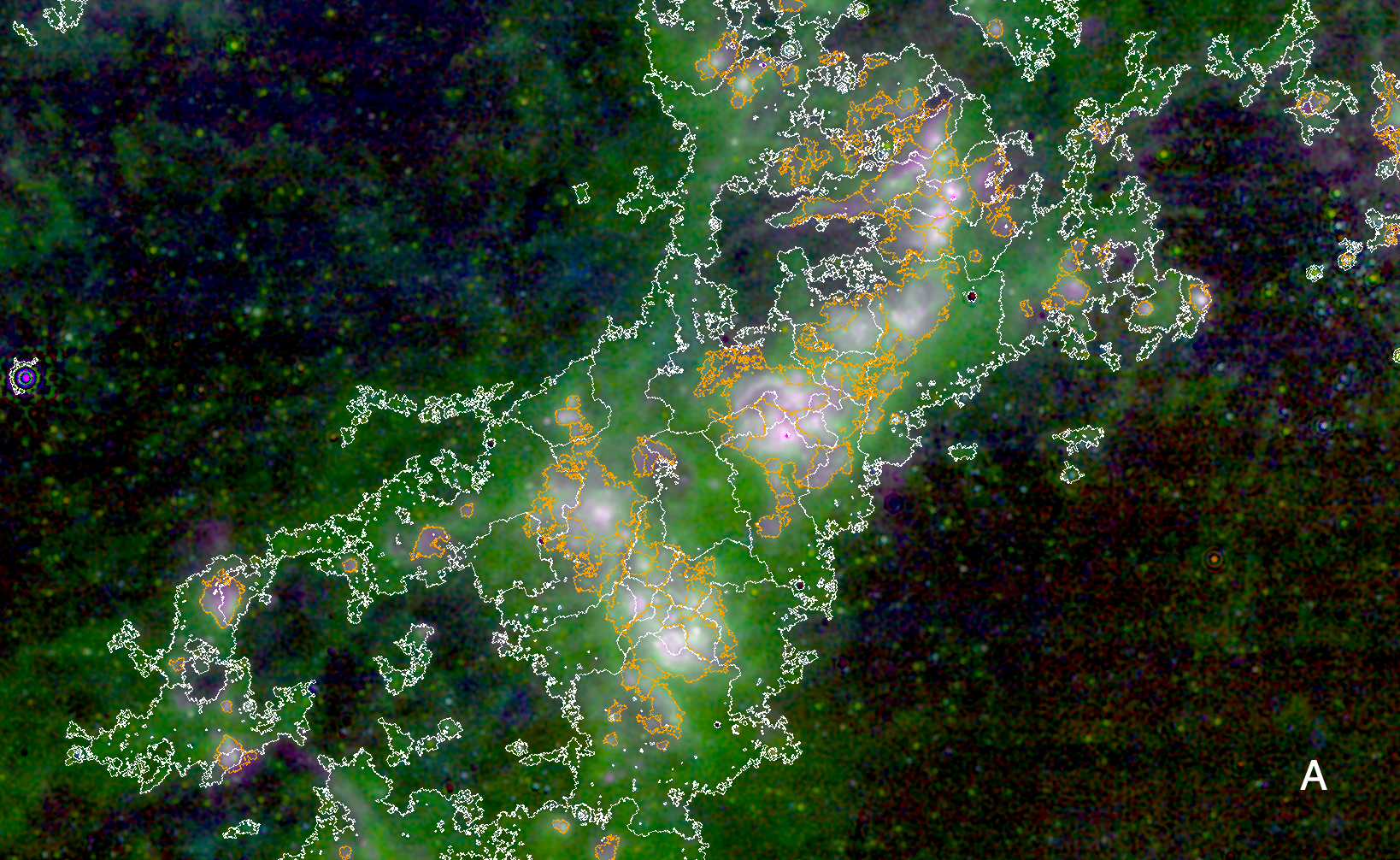}\hfill
    \\[\smallskipamount]
    \includegraphics[width = 0.195\textwidth,trim={4.30cm 1.35cm 4.30cm 1.35cm},clip]{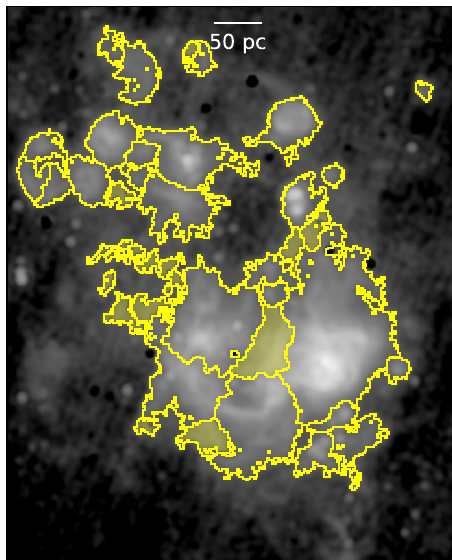}\hfill
    \includegraphics[width = 0.195\textwidth,trim={4.30cm 1.35cm 4.30cm 1.35cm},clip]{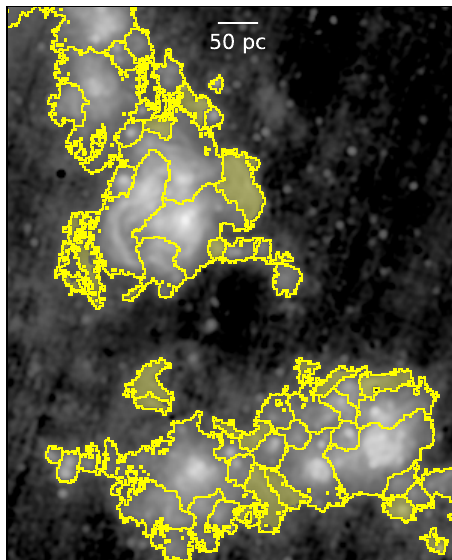}\hfill
    \includegraphics[width = 0.195\textwidth,trim={4.30cm 1.35cm 4.30cm 1.35cm},clip]{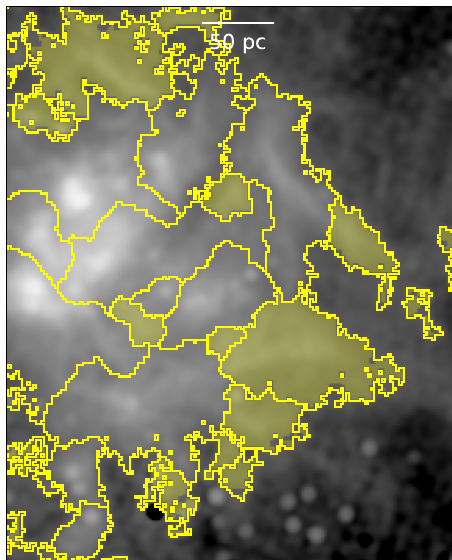}\hfill
    \includegraphics[width = 0.195\textwidth,trim={4.30cm 1.35cm 4.30cm 1.35cm},clip]{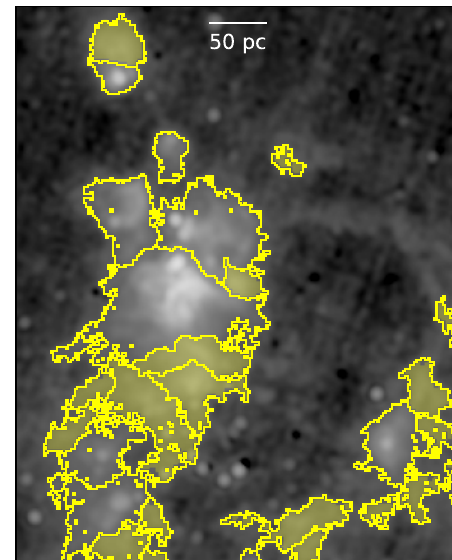}\hfill
    \includegraphics[width = 0.195\textwidth,trim={4.30cm 1.35cm 4.30cm 1.35cm},clip]{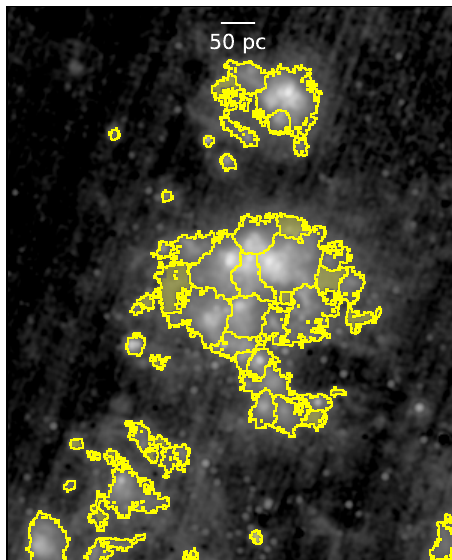}\hfill
  
    \caption{\textit{Top}: RGB JWST image of NGC 628. Blue: Pa$\alpha$ (F187N), green: 3.3 $\mu$m PAH (F335M), red: Br$\alpha$ (F405N). {\it Middle}: RGB zoomed-in views of star-forming regions. These are representative cases of compact and exposed star-forming complexes in the galaxy. Orange (white) contours represent the HII regions (3.3 $\mu$m bright regions) identified using the Pa$\alpha$ (3.3 $\mu$m PAH) map. \textit{Bottom}: Pa$\alpha$ zoomed-in views of star-forming regions. Yellow contours correspond to the identified star-forming regions. The filled regions are identified as diffuse regions and are excluded from our morphological analysis.}
    \label{fig:rgbFULL}
\end{figure*}

JWST/NIRCam and MIRI observations of NGC 628 were obtained in January 2023. An in-depth description of the  data reduction process is presented in Adamo et al.~(in prep.), while we provide here a short summary. We obtained simultaneous observations in four filters in the Short Wavelength (F115W, F150W, F187N, F200W) and Long Wavelength (F277W, F335M, F405N, F444W) NIRCam channels, with a FULLBOX 4TIGHT dither pattern, resulting in a field of view of $\sim 2.2 \arcmin \times 6\arcmin$, covering up to 8 kpc in galactocentric distances (Figure~\ref{fig:rgbFULL}). MIRI observations have been conducted in the F560W and F770W filters. Five MIRI pointings were necessary to cover the same area of the NIRCam mosaic. For the MIRI observations, an external sky background pointing was also included for calibration (RA: 1:36:10.170, Dec: +15:45:54.80). NIRCam data have been reduced with pipeline version 1.12.5 and referencing calibration data context number 1169 and have been obtained from the Mikulski Archive for Space Telescopes (MAST). 
For the MIRI observations, the level 1 (rate.fits) data products for both the target and background fields were downloaded from the MAST. We then used the stage two pipeline to create a master background for each of the MIRI filters, which was then subtracted from each of the corresponding exposures of the target before the remainder of the level 2 processing.  We also replaced the default  SkyMatchStep of the JWST level 3 pipeline with the PixelSkyMatchStep \citep{VarunSoftware}.  The default pipeline sky matching computes background offsets by computing the differences between medians of pixels in overlapping regions of pairs of exposures.  However, this approach often creates strong gradients in the background of the final mosaic for MIRI data. The PixelSkyMatchStep first re-projects the exposures onto the same pixel grid, and computes the median of per pixel differences in overlapping regions (median of differences as opposed to the default difference of medians) and thus removes the background gradients. The remainder of the level 3 processing proceeds as normal and creates a single mosaic for each filter. The image data are then converted from units of MJy/sr to Jy/pixel.  

We also obtained archival HST/ACS F555W, F814W, and F658N (program 9796, PI: Miller, J. and program 10402, PI: Chandar, R.), WCF3/UVIS F275W and F336W (LEGUS program, \citealt{Calzetti15}), WCF3/UVIS F547M and F657N (program 13773, PI: Chandar, R.) images of NGC 628 and we applied standard data reduction steps. Different HST pointings have been combined into mosaics. 

The alignment of all final HST, NIRCam and MIRI mosaics are achieved using GAIA astrometry \citep{Gaia23} to a precision of 15 mas. The NIRCam data and HST data are resampled to the same pixel scale of 0.04\arcsec/px. The MIRI data are resampled to a scale of 0.08\arcsec/px.


The analysis conducted in this manuscript relies on 1.87 $\mu$m-Pa$\alpha$, 4.05 $\mu$m-Br$\alpha$ and 3.3 $\mu$m, 7.7 $\mu$m  PAH emission maps. 
Details of the continuum-subtraction method are described in \cite{Gregg24}, we summarise here the main steps.
To obtain the emission line maps, we subtracted the stellar and dust continuum from the final reduced images in an iterative way. We used both broad and narrow band filters containing the emission features. More specifically, we subtracted the continuum from the F187N narrow filter, which contains the 1.875 $\mu$m-Pa$\alpha$ line, using the F150W and the F200W filters. The F335M and F405N filters contain the 3.3 $\mu$m PAH and the 4.051 $\mu$m-Br$\alpha$, respectively. To subtract the continuum from these two filters, we utilized the F277W and the F444W filters.
Before continuum-subtraction, the data have been convolved to the F444W lowest-resolution point spread function (PSF). The F444W filter has a full-width-at-half-maximum (FWHM) of 0.145$\arcsec$ \citep{Rigby23}.
Regarding the MIRI observations, we convolved the 5.6 $\mu$m PAH emission map to the F770W PSF \citep[FWHM $\approx$ 0.269$\arcsec$,][]{Rigby23}. The 7.7 $\mu$m emission map is derived by subtracting the dust continuum from the F770W image using the F560W. To compare this data with the NIRCam ones, we made additional NIRCam maps convolved to the F770W PSF and we re-gridded them to the MIRI pixel scale.

Each NIRCam continuum-subtracted map is then normalized to 0 by subtracting a 2D plane modeled with the mode values of a set of sky regions selected within the field of view.  Additionally, we estimated the RMS of each map from these regions. We show the set of sky regions utilized for the normalization, for both the 1.87 $\mu$m Pa$\alpha$ and the 3.3 $\mu$m PAH emission features, in the Appendix~\ref{bkgNorm}.

In Fig.~\ref{fig:rgbFULL}, we present an RGB color composite image of the JWST field of view of NGC 628. In blue and red we show the 1.87 $\mu$m Pa$\alpha$ and 4.05 $\mu$m Br$\alpha$ emission line maps. These recombination lines trace gas ionized by massive stars within HII regions. Their emission traces the spiral arms and acts as lamp-posts for massive star formation sites. The 3.3 $\mu$m PAH emission feature is shown in the green channel. This emission is intense in proximity of star-forming regions, tracing the PDRs surrounding HII regions. However, diffuse emission permeates the field of view, revealing the hierarchical structure of the cold ISM.

\section{Methods}
\label{sec:methods}
\subsection{Emerging YSCs catalog}
\label{sec:eyscs_cat}
In Adamo et al.~(in prep.), we present the selection and photometric properties of 3 different classes of eYSCs. This recently formed cluster population is largely missed in UV-optical broad-band selections. They have been detected in the Pa$\alpha$ (Br$\alpha$) and/or 3.3 $\mu$m PAH maps as compact peaks in emission. The first class (eYSCI) contains sources with overlapping peaked emission in both ionized H and PAH, corresponding to the most embedded star clusters in NGC 628. In the second class, the eYSCII, the peaked 3.3 $\mu$m PAH is not present, suggesting a spatial decoupling between PDR and HII region at the location of the eYSC. A third class, compact peaks in the 3.3 $\mu$m PAH, but not in ionized H emission, is associated with star-forming regions, but might indicate regions that do not host ionizing O stars. As shown by mid-IR color properties (Adamo et al., in prep.) and confirmed with spectral energy distribution (SED) analyses (Linden et al., in prep.), eYSCI and II are linked by evolution, while the 3.3 $\mu$m PAH peaks is a mixed class. These 3 classes map the emergence phase of star clusters from their natal clouds. In Linden et al.~(in prep.), we used CIGALE \citep{Boquien19} to fit the observed SED {\rm with HST and NIRCam bands convolved to the F444W PSF} from 0.3 to 5 $\mu$m, estimated by performing aperture photometry using 4 pixel radius (0.16$\arcsec$ $\sim$ 7.6 pc). The derived eYSCs physical properties (age, mass, extinction, PAH abundance, among many others) will be used in the analysis presented in this work.  

\subsection{Identification of star-forming regions}
\label{sec:SFandPDR}

This study is conducted at the highest common spatial resolution achievable with this dataset (0.145$\arcsec$ $\sim 7$ pc with NIRCam at 9.84 Mpc). The goal is to study the morphological and physical properties of star-forming regions as they are transformed by the stellar feedback of their eYSCs. We refer to star-forming regions as entities formed by HII regions (as traced by ionized H emission) and PDRs (as traced by 3.3 $\mu$m PAH emission). As JWST has already started unveiling, PAH features show variations within PDRs that depend on the physical conditions and on the properties of the radiation field \citep{Schroetter24}. For this reason, it is important to note that our definition of PDRs using a single PAH band is simplistic, as a full set of PDR-tracers is required to completely identify PDRs. 
Ionized gas and PAH emissions trace media of different density and temperature ($n_{\rm H} \approx 10^2$ cm$^{-3}$, T$\approx 10^4$ K in the ionized gas and $n_{\rm H} \approx 5-10$ $\times 10^4 $ cm$^{-3}$, T of a few 100s K in PDRs; \citealt{Draine11,Peeters23}) and thus do not spatially coincide. 
Therefore, we independently extracted HII regions and PDRs as surface brightness enhancements in Pa$\alpha$ and 3.3 $\mu$m maps, respectively. The identification proceeded in 3 steps in both tracers.

In the first step, we identified the outer contour of the extended emission with the python package \texttt{ASTRODENDRO}, down to a surface brightness (SB) limit of $2.36 \times 10^{-16}$ erg s$^{-1}$ cm$^{-2}$ arcsec$^{-2}$ and $2.16 \times 10^{-16}$ erg s$^{-1}$ cm$^{-2}$ arcsec$^{-2}$ for the Pa$\alpha$ and the 3.3 $\mu$m PAH continuum subtracted maps, respectively. These limits were achieved by setting a detection level of 5 times the sky rms values (see Sec.~\ref{sec:obs&dr}) and a required minimum number of pixels for an independent region of 40 and 100 for the Pa$\alpha$ and the 3.3 $\mu$m PAH emissions, respectively. These areas correspond to 0.064 arcsec$^2$ $\sim$ 145 pc$^2$ and 0.16 arcsec$^2$ $\sim$ 364 pc$^2$, or circularised radii of 7 ($\sim$0.147$\arcsec$) and 11 pc ($\sim$0.230$\arcsec$), respectively. 
Because of strong residuals due to the bulge light, we masked the central region of the galaxy with a circle of 152 pixels radius (6.08$\arcsec$ $\sim$ 0.29 kpc).
The software \texttt{ASTRODENDRO} computed a dendrogram of outer contours. For the scope of this study, we only identified contours that correspond to the outermost trunk, without taking into account sub-structures.

In the second step, we used a multi-thresholding technique from the \texttt{segmentation.deblend\_sources} method in the \texttt{PHOTUTILS} python package \citep{Bradley23}, to deblend the large complexes extracted by \texttt{ASTRODENDRO}.
This method generated new segmentation images allowing us to break down HII regions and PDRs traced by 3.3 $\mu$m emission at the desired resolution. After several tests, visual inspection and comparisons with the point source extraction performed in Adamo et al.~(in prep.), we established the segmentation parameters to: 40 multi-thresholding levels, 5 $\times 10^{-4}$ as value of contrast (the fraction of the flux needed for a local peak to be identified as a single object), and 30 minimum number of pixels, where the latter corresponds to a circular radius of about 3.1 pixels (0.12$\arcsec$ $\sim$ 6 pc). We present zoomed-in views of the identified HII regions and PDRs traced by 3.3 $\mu$m emission in the middle panels of Fig.~\ref{fig:rgbFULL}. Orange contours are extracted from the Pa$\alpha$ map, while white contours are from the PAH emission, revealing the more extended and diffuse nature of the PDRs. However, our segmentation procedure simply breaks down regions based on surface brightness contrast that are not necessarily physically motivated. Therefore, a third final step was required to distinguish regions containing powering stellar sources from simply diffuse patches of emitting gas.

In this final step, we used the catalog of eYSCI and eYSCII to discriminate between HII regions and diffuse emission identified in the Pa$\alpha$ map.
More specifically, we flagged a segment as diffuse emission when it is not associated to one or more eYSCs. We considered a cluster being associated to a region if the area of a circular aperture of 4 pixels (0.16$\arcsec$ $\sim$ 7.6 pc) from the position of the cluster overlaps with the area of the region delimited by the contours.
Analogously, we distinguished between compact and diffuse emission in the 3.3 $\mu$m PAH map looking for overlaps with eYSCI and 3.3 $\mu$m PAH peaks along the line of sight. We show examples of this discrimination for the Pa$\alpha$ emission in the bottom panels of Fig.~\ref{fig:rgbFULL}, where the regions tagged as diffuse emission are highlighted as yellow-filled contours. As our analysis is based on a statistical sample, the influence of geometric effects, is expected to be mitigated through averaging.

The initial \texttt{ASTRODENDRO} extraction resulted in 996 and 1532 regions in the Pa$\alpha$ and 3.3 $\mu$m maps. After applying the segmentation deblending step, their number increased to 1708 and 7127 regions, respectively. Matching with physically informed sources (eYSCs) resulted in 962 HII regions and 1084 PDRs traced by PAH emission detected in the Pa$\alpha$ and 3.3 $\mu$m maps, respectively.

\begin{deluxetable*}{ccccc}

\tablecaption{HII regions catalog extracted by means of the Pa$\alpha$ emission line map \label{tab:catHII}}

\tablenum{1}

\tablehead{\colhead{ID} & \colhead{ra} & \colhead{dec} & \colhead{Area [arcsec$^2$]} & \colhead{F$_\nu$ [Jy]} 
 }

\startdata
1 & 24.184746 & 15.734762 & 0.5040 & 2.979e-05 \\
2 & 24.184608 & 15.734973 & 0.1632 & 6.884e-06 \\
6 & 24.187228 & 15.736117 & 0.4384 & 8.159e-06 \\
7 & 24.177105 & 15.736251 & 0.1088 & 3.412e-06 \\
8 & 24.174865 & 15.736318 & 0.3888 & 8.247e-06 \\
12 & 24.175050 & 15.737329 & 0.4576 & 1.040e-05 \\
16 & 24.193866 & 15.739150 & 0.5088 & 1.042e-05 \\
18 & 24.189653 & 15.739717 & 0.3968 & 8.949e-06 \\
19 & 24.191361 & 15.740084 & 0.5440 & 2.134e-05 \\
20 & 24.180568 & 15.740118 & 0.0720 & 2.107e-06 \\
\enddata

\tablecomments{Table 1 is published in its entirety in the machine-readable format.
A portion is shown here for guidance regarding its form and content.}
\end{deluxetable*} 

\begin{deluxetable*}{ccccccc}
\tablecaption{Catalog of PDRs extracted by means of the 3.3 $\mu$m emission map \label{tab:catPAH}}

\tablenum{2}

\tablehead{\colhead{ID} & \colhead{ra} & \colhead{dec} & \colhead{Area [arcsec$^2$]} & \colhead{F$_\nu$ [Jy]} 
 }

\startdata
3 & 24.172903 & 15.724875 & 0.3184 & 8.596e-07 \\
4 & 24.172461 & 15.724893 & 0.5056 & 1.708e-06 \\
8 & 24.178690 & 15.725939 & 0.3056 & 9.929e-07 \\
20 & 24.174753 & 15.728091 & 1.1296 & 3.650e-06 \\
32 & 24.175551 & 15.730967 & 0.4960 & 2.076e-06 \\
37 & 24.190390 & 15.731289 & 0.8672 & 2.309e-06 \\
49 & 24.196100 & 15.732951 & 0.2032 & 1.131e-06 \\
68 & 24.196027 & 15.735095 & 0.3776 & 1.663e-06 \\
89 & 24.171391 & 15.737656 & 0.3136 & 9.776e-07 \\
144 & 24.192023 & 15.744615 & 0.1680 & 5.629e-07 \\
\enddata

\tablecomments{Table 2 is published in its entirety in the machine-readable format.
A portion is shown here for guidance regarding its form and content.}
\end{deluxetable*}

The catalogs of HII regions and 3.3 $\mu$m extracted PDRs are presented in Tab.~\ref{tab:catHII} and Tab.~\ref{tab:catPAH} and contain the coordinates of the center, the area and the flux emitted within the segment. The circularised radius is defined from the area $A$ of each region: R$_{\rm C} = \sqrt{A/\pi}$. We adopt the position of the brightest pixel in the Pa$\alpha$ map as the center of the HII region for simplicity, assuming the peak of the ionized gas emission directly traces the position of the massive stars responsible to produce ionizing photons.
For the PDRs traced by 3.3 $\mu$m PAH catalog, the center of each region corresponds to the first flux moment of the pixels within the respective segment.

The Pa$\alpha$ flux contained within the extracted HII regions represents 90\% of the total emission detected above the 5$\sigma$ surface brightness limit listed above. On the other hand, only 60\% of the 3.3 $\mu$m PAH emission is contained in the identified compact 3.3 $\mu$m bright regions, evidencing the more diffuse nature of the PAH emission.

\subsection{Properties of HII regions and PDRs traced by 3.3 $\mu$m PAH emission as a function of galactocentric distances} 
\label{sec:methods:properties}
\begin{figure*}[htb]
    \centering
    \includegraphics[width = \textwidth]{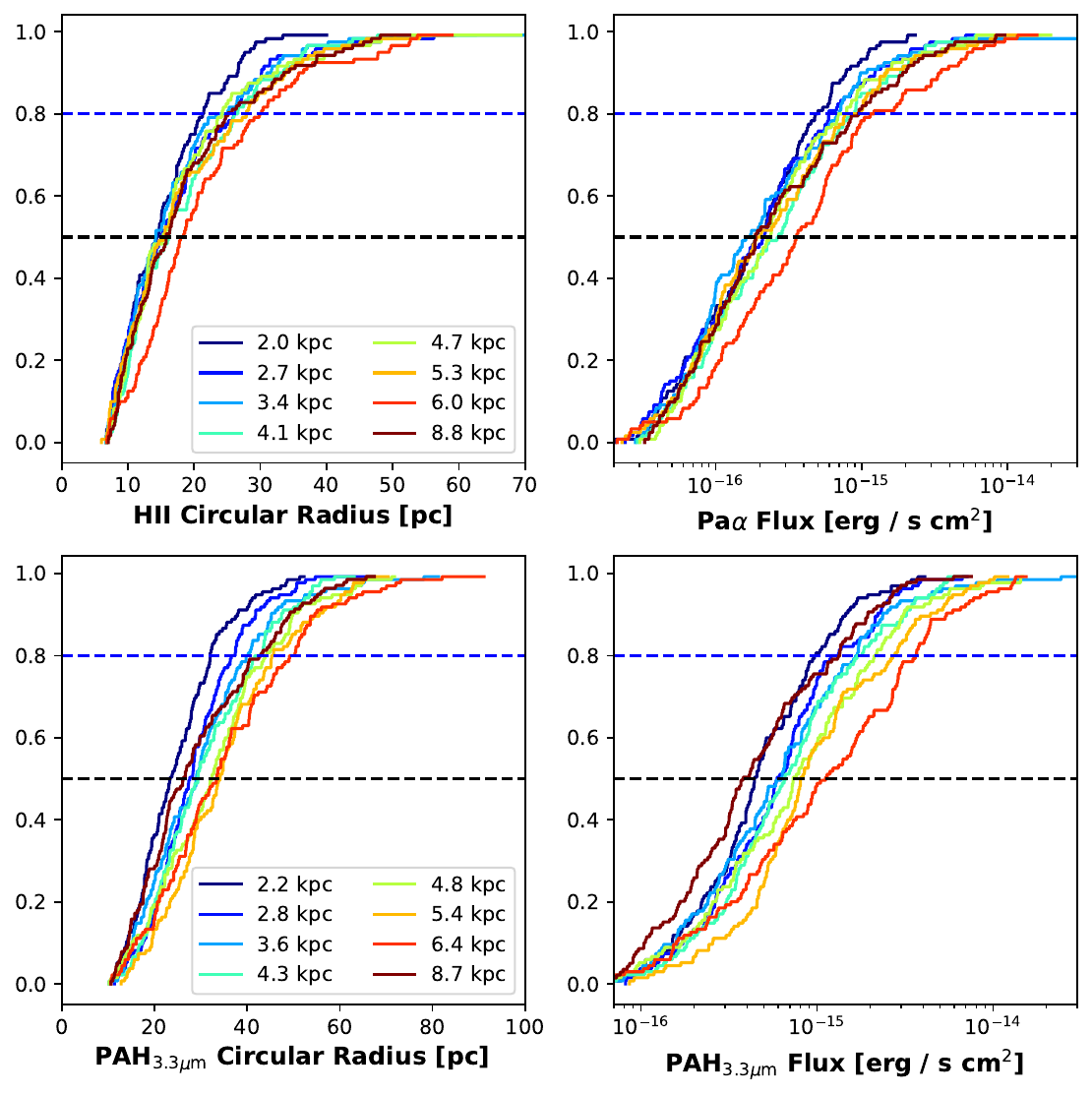}
    \caption{Cumulative distributions (\textit{top:} Pa$\alpha$, \textit{bottom:} 3.3 $\mu$m) of both sizes (left) and fluxes (right) in our sample of compact regions binned for different galactocentric distances. Legends show the distance from the center of the galaxy at the outer edge of each bin. Each bin contains the same number of regions (130 HII regions and 150  3.3 $\mu$m PAH bright regions per bin)}.  Black and blue dashed lines correspond to cumulative fractions $f = 0.5$ and $f = 0.8$, respectively.
    \label{fig:galdist}
\end{figure*}

We investigate how fluxes and sizes of HII regions and 3.3 $\mu$m PAH bright regions correlate with the radial distance by dividing the maps into 8 radial bins.
We constrain the size of the bins by requiring that each bin contains an equal number of regions (130 HII regions and 150 3.3 $\mu$m PAH bright regions per bin), ensuring uniform density and preventing contamination in the analysis resulting only from differences in the number of regions per bin. The legend of Fig.~\ref{fig:galdist} lists the distance of the outer edge of each bin for Pa$\alpha$ and 3.3 $ \mu$m PAH emission.

We present the cumulative distributions of the circular projected radii and the total flux in each segment for Pa$\alpha$ and 3.3 $\mu$m PAH emission in Fig.~\ref{fig:galdist}. Focusing on the HII regions (top panels of the figure), we observe more compact and less luminous regions in the innermost parts of the galaxy. Moving outward from the center of the galaxy we notice more extended and brighter regions, especially for regions located in bin 7 (5.3 - 6 kpc). The region of the disk corresponding to this bin also reveals a prominent CO(1 $\rightarrow$ 0) emission and coincides with the co-rotation radius of the spiral pattern \citep{Cepa90,Rebolledo15, Herrera20}. This feature is observable as a region of intense star-formation at about 6 kpc from the center of the galaxy (region {\it A} of Fig.~\ref{fig:rgbFULL} and surroundings). Considering the 3.3 $\mu$m map, we observe that PAH emission traces significantly larger regions which are more extended and brighter with increasing galactocentric distance, similarly to the HII regions. These observed trends disappear in the last bin (bin 8, 6 - 8.8 (6.4 - 8.7) kpc). 
The brightness of the 3.3 $\mu$m-regions increases with radius by a factor of $\sim 2-3$ (about one magnitude). \cite{Elmegreen19} analyzed 8$\mu$m concentrations with Spitzer at poorer spatial resolution ($\sim$ ten times lower), finding a slight  decreasing brightness at increasing galaxy radii. This observed discrepancy could be explained by the different physical scales explored in the two works.

Thanks to the high spatial resolution provided by these data, we can now resolve HII regions at much smaller scales than obtained with ground-based observations. In the distribution of sizes (circular radii), we find a value of $\approx 15$ pc for the 50\% percentile, a factor of $\sim 2$ smaller than the HII regions catalog extracted with MUSE for this galaxy \citep[35 pc,][]{Groves23}. In all the defined bins, at least 50\% of the HII regions have sizes smaller than 20 pc. On the other hand, PDRs traced by 3.3 $\mu$m PAH emission show a broader distribution of sizes, being on average two times larger than HII regions. We note that even though we minimized the user dependency of the identification process by using different sets of parameters, these results may still be slightly affected by the choice of parameters.

\subsection{PAH emission in star-forming regions}
\label{sec:PAHinSF}

As already mentioned, the morphological evolution of PDRs in star-forming regions is shaped by the stellar feedback from eYSCs, which modify the content and the distribution of PAH molecules, direct tracers for PDRs. To characterize the distribution of PAH emission, we constructed surface brightness profiles of the star-forming regions. In this analysis, we use as reference the catalog of HII regions. Therefore, by definition, we focus on star-forming regions that contain at least one compact HII region with one or more eYSCs. 
We removed nine regions due to over/under-subtraction of the stellar continuum, ending up with 953 regions for our analysis. 
For each region and emission map, we employed circular annulus apertures to perform aperture photometry (\texttt{PHOTUTILS}) centered on the brightest Pa$\alpha$ peak, from 1 pixel up to $2 \times {\rm R}_{\rm C}$, with steps of 3.5 pixels (0.14$\arcsec$ $\sim$ 6.68 pc $\sim$ FWHM of the F444W PSF) between each aperture. The first aperture is performed using a circular aperture of 1 pixel radius. The choice of the upper limit in the circular annulus apertures ($2 \times {\rm R}_{\rm C}$) is dictated by the fact that 3.3 $\mu$m bright regions are more extended (on average a factor of two) than the HII regions, as explained in Sec.~\ref{sec:methods:properties}.
To improve the flux estimation, we corrected for local background using the mode value of an annulus aperture of $R_{\rm min} = 50$ pixels and $R_{\rm max} = 55$ pixels. The chosen background annulus apertures stand well outside the star-forming regions, since the median and the maximum values of the radii in our sample of star-forming regions are $\approx$ 8 and $\approx$ 45 pixels ($\sim 15$ and $\sim 85$ pc), respectively. To reduce contamination from adjacent regions we applied sigma clipping and used the mode of the background flux within the annulus. We note that the recovered trends do not change if the local sky background is not applied.
 
\begin{figure*}[htb!]
    \centering
    \includegraphics[width = 0.496\textwidth]{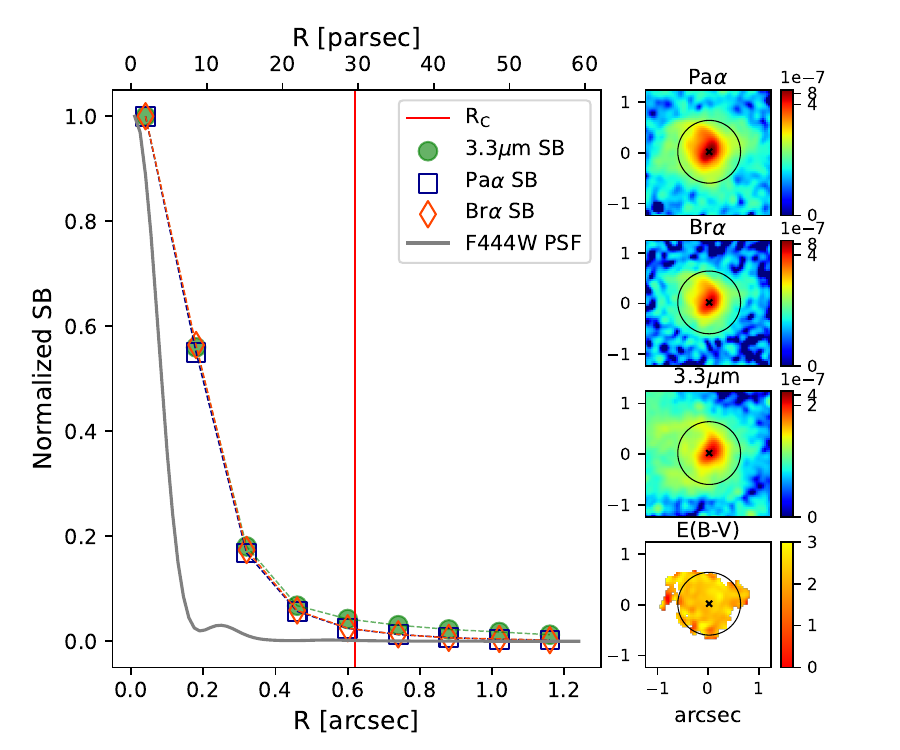}
    \hfill
    \includegraphics[width = 0.496\textwidth]{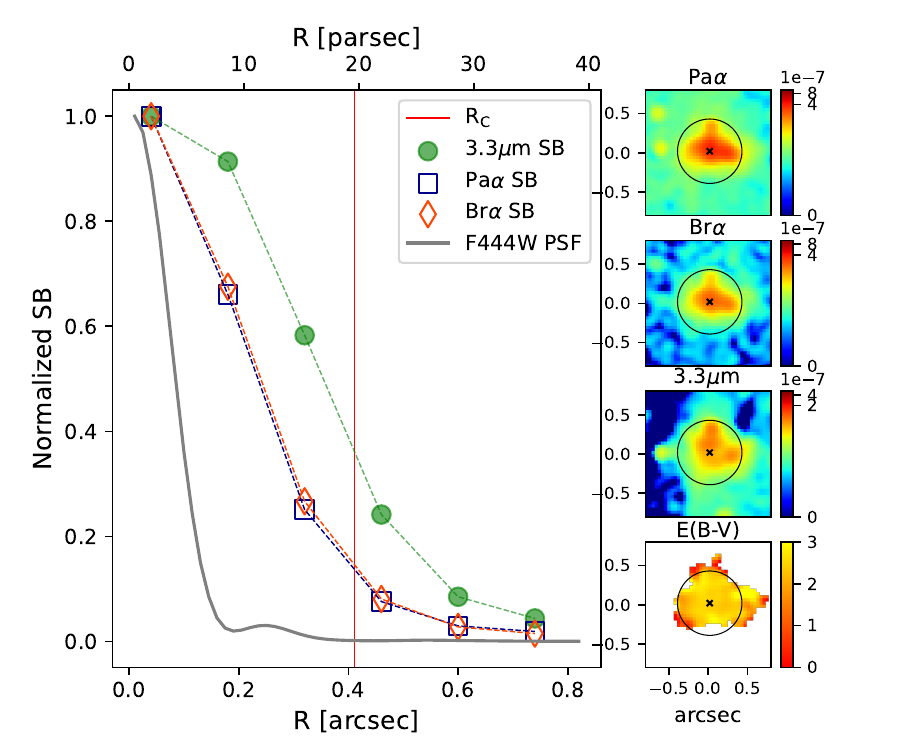}
    \\[\smallskipamount]
\includegraphics[width = 0.496\textwidth]{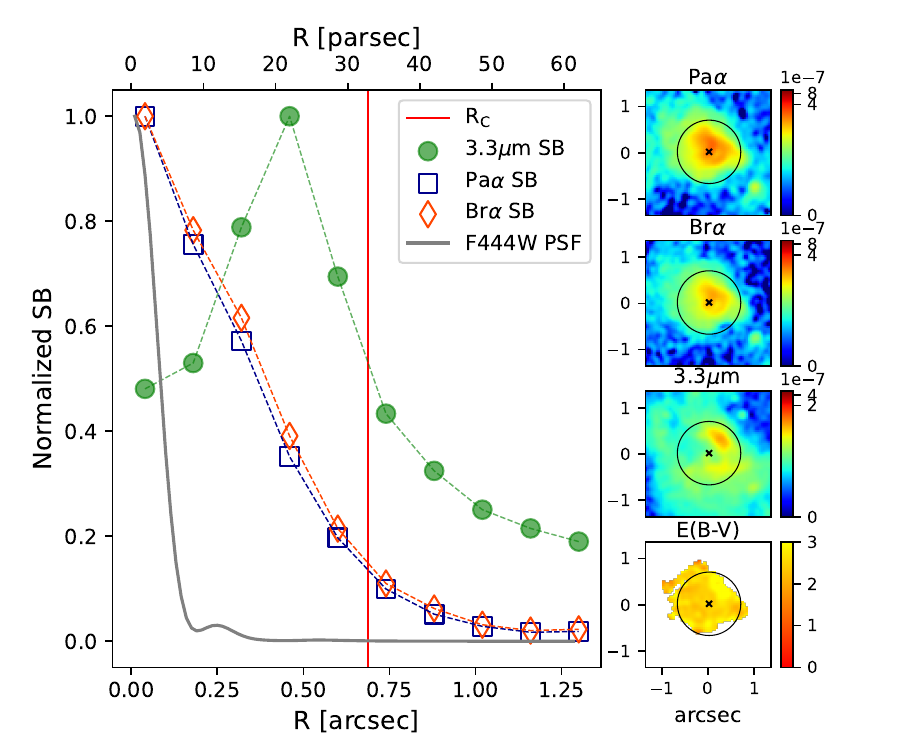}

    \caption{Normalized SB profiles for three different star-forming regions. The top left one represents an example of compact PAH, while the top right and the bottom ones represent examples of an extended and an open region, respectively. For each region, the left panel displays the SB profiles of the Pa$\alpha$ emission (dark blue un-filled squares), the 3.3 $\mu$m PAH emission (green filled circles) and the Br$\alpha$ (red un-filled diamonds). The grey line represents the normalized SB profile of the PSF in the F444W filter. 
    Red vertical solid lines represent the circular radius R$_{\rm C}$ of each region. The four right panels show a zoom-in in each region in the Pa$\alpha$, Br$\alpha$, 3.3 $\mu$m PAH continuum subtracted frames (log scale, units are Jy), as well as the E(B-V) map in the bottom panel. White pixels in the E(B-V) map correspond to the area outside the Pa$\alpha$ SB selected contours of the respective HII region. Black crosses are in the center of the region, corresponding to the brightest pixel in the Pa$\alpha$ map. For each panel, the black circle is centered in the center of the region, with a radius of R$_C$.
    }
    \label{fig:PAHinSF}
\end{figure*}

Finally, for each measurement, we divided the observed flux by the respective area and then we normalized it by the maximum value, to obtain normalized surface brightness profiles. In Fig.~\ref{fig:PAHinSF} we show examples of the resulting profiles.

\begin{figure*}[htb]
    \centering

    \includegraphics[width =\textwidth]{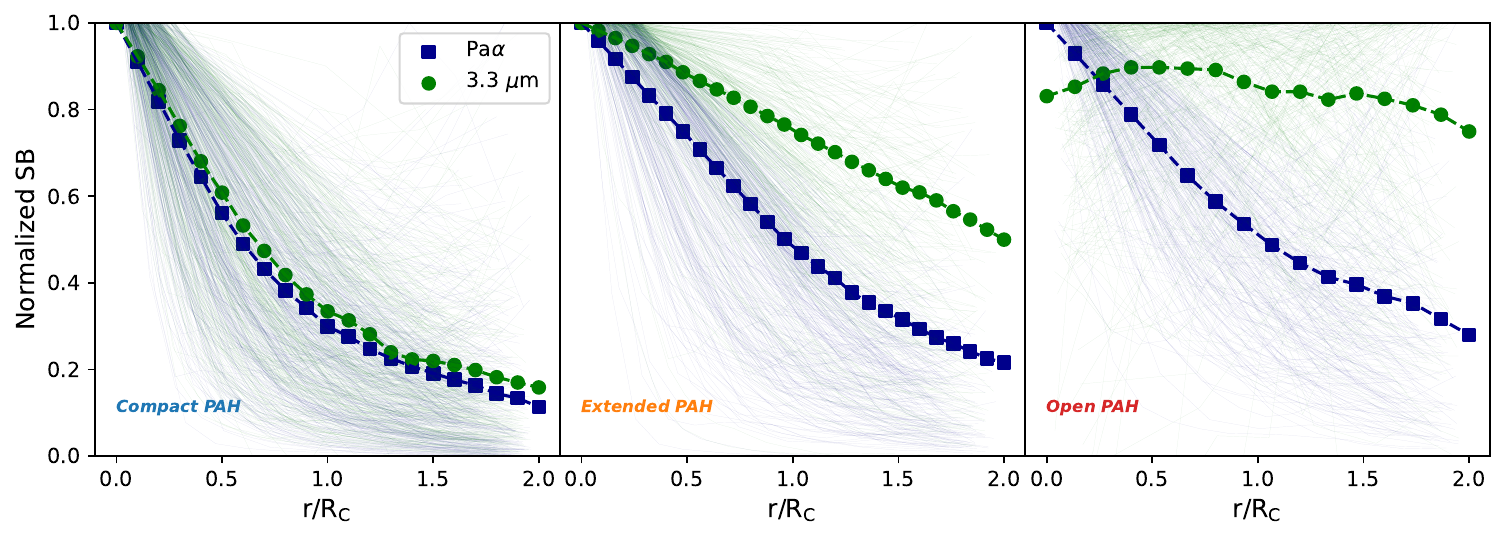}\hfill
    \\[\smallskipamount]
    \includegraphics[width = \textwidth]{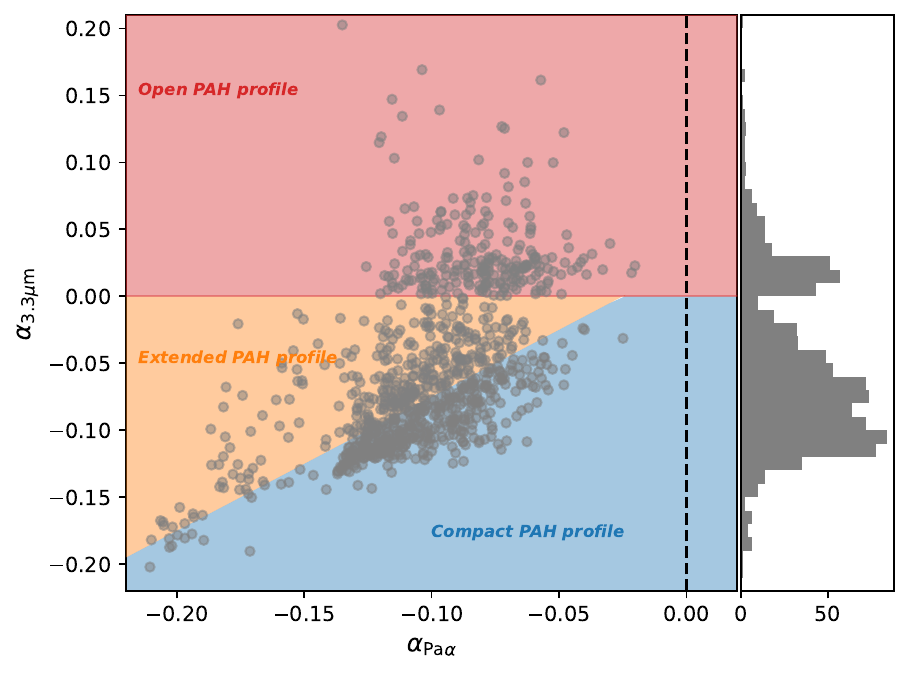}
    \caption{Morphological classification of the PAH emission in star-forming regions. {\it Top:} Normalized SB profiles of Pa$\alpha$ and 3.3 $\mu$m emission for the entire sample of star-forming regions in the three different morphologies. Blue filled squares and green filled circles represent the Pa$\alpha$ and 3.3 $\mu$m median SB profiles, respectively. For each PAH morphology, we obtained the median profiles by interpolating the SB profiles to the maximum number of radial bins defined by the larger region in the sample. 
    {\it Bottom:} the scatter plot shows the relationship between $\alpha_{3.3 \mu{\rm m}}$ and $\alpha_{{\rm Pa}\alpha}$ in the entire sample of star-forming regions, where $\alpha$ is the slope of the normalized SB profile for the emission in each region, as defined in the text. By construction, $\alpha_{{\rm Pa}\alpha}$ is always negative, while the slopes defined by the PAH SB emission profiles span a broader range, from negative to positive values, which defines the three morphological classes of the 3.3 $\mu$m PDRs. The vertical dashed black line is the upper limit for $\alpha_{{\rm Pa}\alpha}$ by construction of the SB profiles. Different background colors in the plot represent the morphological classification (compact, extended and open) of the PAH emission, obtained by a visual inspection of the morphology of the regions to set boundaries in the $\alpha_{3.3 \mu{\rm m}}$- $\alpha_{{\rm Pa}\alpha}$ space.
    }
    \label{fig:morphoClass}
\end{figure*}

In these profiles, the maximum of the Pa$\alpha$ and Br$\alpha$ SB is always in the first bin by construction. On the other hand, the radial distribution of the PAH SB shows different behaviours.
Specifically, we distinguish three types of PAH SB profiles:
\begin{itemize}
    \item[--] {\bf Compact PAH profile}: the PAH emission follows the H recombination line emission (top left panel of Fig.~\ref{fig:PAHinSF}).
    \item[--] {\bf Extended PAH profile}: the PAH profile is more diffuse and extended compared to the Pa$\alpha$ and Br$\alpha$ SB profiles (top right panel of Fig.~\ref{fig:PAHinSF}).
    \item[--] {\bf Open PAH profile}: the peak of the PAH SB profile lies away from the center, and visually the region has a \emph{donut}-like appearance, with a notable dearth of PAH in the central region (third panel in Fig.~\ref{fig:PAHinSF}).
\end{itemize}
For a high number of regions, the Pa$\alpha$ to Br$\alpha$ ratio is consistent with low and sometimes negatives values of E(B-V) attenuation. We have investigated this issue and we tend to attribute it to an underestimation of the Br$\alpha$ flux. We report the results of this investigation in the Appendix~\ref{attenuation} and leave analyses of attenuation to future studies.

To distribute the regions in the three classes, we derived the slope of the SB profiles in the central pixels, linearly fitting the normalized SB profiles for both the Pa$\alpha$ and the PAH emissions. 
For PAH profiles where the peak of the SB profiles is in the center, we employed the linear fit on the first three radial bins to both Pa$\alpha$ and PAH, corresponding to 10.5 pixels (0.42$\arcsec$ $\sim$ 20 pc) in size for both emissions. On the other hand, if the PAH peak lies outside the center, we linearly fitted both the Pa$\alpha$ and PAH profiles up to the peak of the PAH SB profile. For the smallest regions, we fitted only two radial bins.
\\
From the linear fit, we extract the values of the slopes in the two tracers, obtaining $\alpha_{{\rm Pa}\alpha}$ and $\alpha_{3.3 \mu{\rm m}}$. The distribution of the slopes is plotted in Fig.~\ref{fig:morphoClass}. Focusing on the distribution, we can immediately notice that, by construction, the Pa$\alpha$ profiles have all negative slopes, indicating declining profiles. 
On the other hand, the PAH emission shows a broad range of slopes that goes from negative values to positive ones. 
By visually examining the regions, we set boundary conditions in the $\alpha_{{\rm Pa}\alpha}$-$\alpha_{3.3 \mu{\rm m}}$ space, that delineate the three morphological types of PAH SB profiles described above. 
The blue-filled area in Fig.~\ref{fig:morphoClass} contains the compact PAH profiles. The orange one contains the extended region and lastly, open profiles lie in the red-filled area. We used the slope-based morphological separation to classify star-forming regions. We note that the distinction between compact and extended profiles is somewhat less defined, with the extended morphologies simply being a transition phase between compact and open ones. Among the 953 star-forming regions in our sample, we identified 401 compact PAH regions ($\sim$ 42\%), 319 extended regions ($\sim$ 34\%) and 233 open regions ($\sim$ 24\%). In the top panel of Fig.~\ref{fig:morphoClass}, we show the normalized SB profiles divided into the three types, showing the overall gradual change of PAH appearance around HII regions.

\subsection{Emerging YSCs}
\label{sec:eyscs_connection}
We link our star-forming regions with their associated eYSC using the catalogs of eYSCs from Adamo et al.~(in prep.), following a similar methodology to what was presented in \cite{DellaBruna22} for the HII region population in M83.
The three classes of eYSCs (eYSCI, eYSCII and 3.3 $\mu$m PAH peaks, defined in Sec.~\ref{sec:eyscs_cat}) represent 2207 objects in NGC 628. For each region, we selected only the cluster located closest to the center of the region and we removed a few more regions where the eYSC is very near the edge of the identified star-forming region, obtaining a sample of 947 star-forming regions with a coinciding eYSC. We are aware of the fact that considering only the central cluster is a simplification. However, we notice that 80-90\% of the star-forming regions contain either one or two clusters and the results in the following analysis do not change whether we consider multiple clusters associated with the same region.

We find that the morphological classification presented in Sec.~\ref{sec:PAHinSF} independently correlates with the eYSC classes. We summarize the relative numbers of eYSCs with respect to the morphological appearance of the PDRs traced by 3.3 $\mu$m PAH in Tab.~\ref{tab:numbers}.
Open regions primarily host (85.5\%) clusters identified as eYSCII, while we predominantly find that eYSCI are in compact regions (77\%). Extended regions host almost equal numbers of eYSCI and eYSCII clusters, suggesting that the boundaries of this sub-sample are not well defined, and that PDRs in these regions are transitioning from compact to open.
Because the center of each star-forming region lies at the center of the HII region, by construction we find only 5 regions coinciding with 3.3 $\mu$m peaks. Visual inspection of these regions shows extended Pa$\alpha$ emission, explaining why they are not identified as eYSCI.

\begin{deluxetable}{lccc}

\tablecaption{Number of compact, extended and open regions identified as either eYSCI, eYSCII or 3.3 $\mu$m peak \label{tab:numbers}}

\tablenum{3}

\tablehead{\colhead{eYSC type} & \colhead{Compact} & \colhead{Extended} & \colhead{Open} 
 } 

\startdata
I & 308 (77\%) & 171 (53\%) & 32 (14\%) \\
II & 89 (22\%) & 146 (46\%) & 196 (85.5\%) \\
3.3 $\mu$m PAH & 2 (1\%) & 2 (1\%) & 1 (0.5\%) \\
\enddata

\end{deluxetable}


\section{Results}
\label{sec:res}
In the previous section, we have presented the 3.3 $\mu$m PAH SB profiles exhibiting various morphologies. Particularly, we have distinguished between compact, extended and open star-forming regions, where this morphological classification relates to the 3.3 $\mu$m PAH emission as a tracer of PDR (see Sec.~\ref{sec:SFandPDR}). 
Additionally, we have spatially linked star-forming regions to their associated central eYSCs. In the following analysis, we assume that the latter stellar population is likely the source of feedback operating in the regions. 
\subsection{Demographics of the eYSCs associated to the star-forming regions}
\begin{figure*}[htb]
    \centering
    \includegraphics[width = 1.0\textwidth]{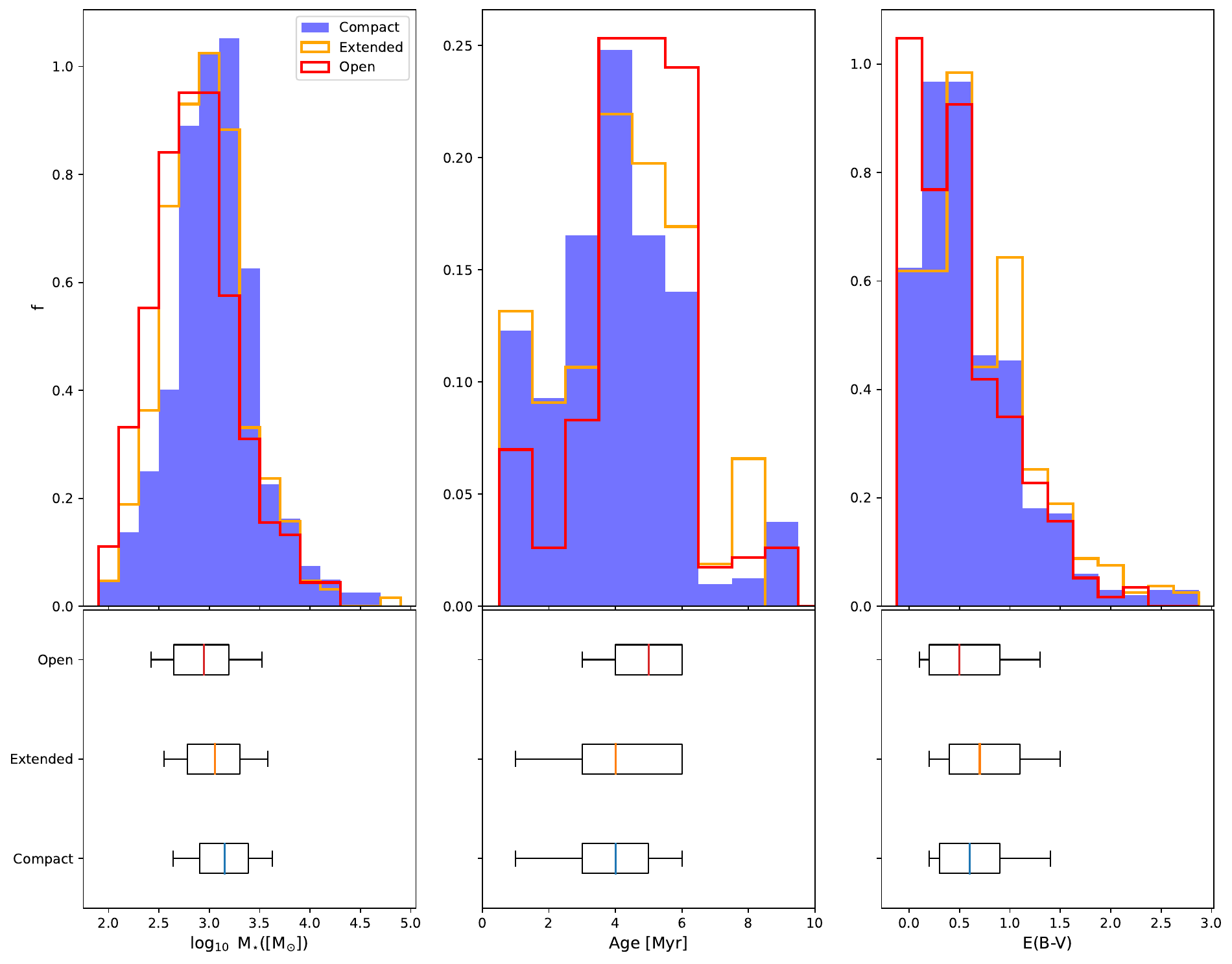}
    \caption{{\it Left}: normalized stellar mass distributions of our sample of eYSCs color-coded by the associated PAH morphology as in Fig.~\ref{fig:morphoClass} (\textit{top}) and relative boxplots of the distributions (\textit{bottom}). Each boxplot extends from the first to the third quartile and the whiskers cover the data-set from the 10th to the 90th percentile. Median values for each distribution are shown as a vertical line in the respective boxplot color-coded by the morphology classification. 
    {\it Middle}: Same as above, but for the age distributions.
    {\it Right}: Same as above, but for the nebular E(B-V) distributions. These properties are derived from the SED fitting from CIGALE (Linden et al., in prep.).}
    \label{fig:massAge_CIGALE}
\end{figure*}
\begin{figure}[htb]
    \centering
    \includegraphics[width = 1.0\columnwidth]{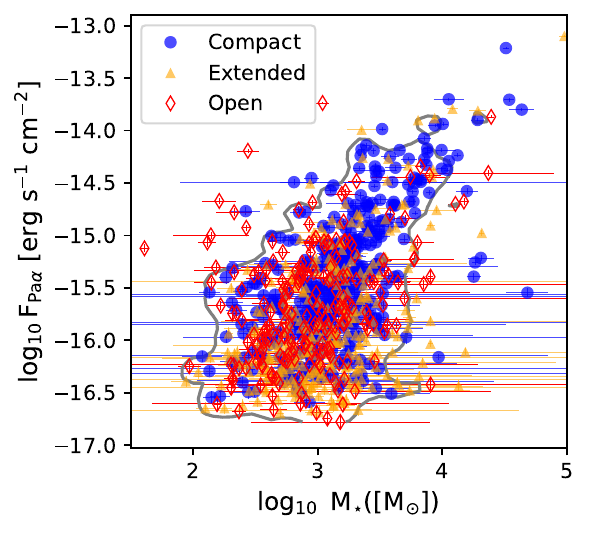}
    \caption{Integrated Pa$\alpha$ fluxes of our sample of star-forming regions as a function of the stellar mass of the linked eYSCs derived from SED fitting from CIGALE. Error bars for the stellar masses correspond to the 68\% confidence interval. The grey solid line represents an iso-density contour after smoothing the data with a Gaussian 2D kernel. Regions within the contour indicates areas where the density of points is above 6\% of the maximum density.
    Data are color-coded by the defined morphology as in Fig.~\ref{fig:morphoClass}.}
    \label{fig:FluxVsMass_CIGALE}
\end{figure}
In Fig.~\ref{fig:massAge_CIGALE} we present the best-$\chi^2$ results of the CIGALE fitting code for the stellar mass (left), the age (centre) and the nebular attenuation (right) of our sample of eYSCs from Linden et al.~(in prep.). 
Different histograms represent the different morphologies of the PDRs traced by 3.3 $\mu$m PAH emission associated to the eYSCs. 
In the bottom panels, the boxplots provide a better view of the covered physical ranges in the different classes.
Focusing on the stellar mass, the boxplot indicates that the mass of the clusters shows weak correlations with the morphology of the 3.3 $\mu$m emission. In general, the distribution of masses aligns with the findings from \cite{Hassani23}, who investigated compact sources identified at 21 $\mu$m in a sample of four nearby galaxies, among them NGC 628. Median values for the three classes range between $\log_{10}$M$_{\star}$([M$_{\odot}$]) $\approx 2.9$ and $\log_{10}$M$_{\star}$([M$_{\odot}$]) $\approx 3.2$. We see that there is slight evidence of more massive eYSCs in the compact class, however, the sample size may be a factor (401 eYSC associated with compact morphology, versus 319 extended, and 233 open classes). 
On the other hand, the age distribution in the middle panels of Fig.~\ref{fig:massAge_CIGALE} exhibits a more significant trend with PAH morphology. The median values associated with compact and extended PAH morphologies are 1 Myr younger than the open counterparts. The majority (75\%) of the compact regions host eYSCs younger than 5 Myr, while the sources associated with the extended class span the whole age range covered by eYSCs. 90\% of eYSCs associated with open PAH morphologies are older than 3 Myr and 75 \% older than 4 Myr. The gradual age gradient, from compact to open, suggests  that the change in morphology of the PDRs traced by 3.3 $\mu$m emission is associated to the temporal evolution of the star-forming regions. 
In the right panels of Fig.~\ref{fig:massAge_CIGALE}, we present the distribution of nebular E(B-V) values, computed with CIGALE, using emission lines and assuming a dust screen geometry \citep[modified starburst attenuation law,][]{Calzetti00}.
We do not see a clear relation among the three morphologies. Median values lie between 0.5 mag for the open class to 0.7 mag for the compact one and 80\% of the eYSCs in the three classes present values of attenuation between 0.1 and 1.5 mag.


In Fig.~\ref{fig:FluxVsMass_CIGALE}, we investigate how the mass derived with SED fitting correlates with the total integrated Pa$\alpha$ flux emitted by the associated star-forming regions. These two quantities are independently estimated and we expect them to correlate. 
The Pa$\alpha$ flux is a first order indicator of the number of ionizing photons available in the region, which scales with the number of massive stars under the condition of an idealized ionization bound HII region with a fully sampled initial mass function (IMF). The CIGALE stellar population synthesis models do not account for stochastic IMF sampling, which dominates when the total mass of the star clusters is about a few hundred solar masses \citep{Calzetti10}. We discuss in Linden et al.~(in prep.), that the presence of emission lines in the observed SEDs somewhat mitigates the missclassification of cluster ages and therefore masses. However, the number of ionizing photons in CIGALE is produced by a rescaling of a fully sampled IMF, therefore, it does not take into account the variable number of massive stars populating low mass clusters.  
When comparing the two quantities, we find a correlation between the mass of the clusters $\log_{10}$M$_{\star}$ and the integrated Pa$\alpha$ flux of the associated HII region. However, we see over two orders of magnitude dispersion in Pa$\alpha$ flux as a function of cluster stellar mass, for masses below a few 1000 M$_{\odot}$.  We interpret this increase in dispersion at the low mass end as an indicator of stochastic sampling of the IMF. Finally, we do not see significant differences among the three morphological classes, but a trend where eYSCs associated with compact regions reach higher masses and luminosities.
As shown by the iso-density contour in Fig.~\ref{fig:FluxVsMass_CIGALE}, the observed dispersion goes down for the compact class, relative to the other two morphologies, suggesting that for the brightest and compact sources, CIGALE recovers the expected trends.
We find no correlation in the Pa$\alpha$ flux vs the age of the cluster, even though we notice slightly higher fluxes for the youngest clusters.

\subsection{PAH band ratio}

To probe PAH emission and variation across  the various morphologies, we investigated the relation between the 3.3 $\mu$m and the 7.7 $\mu$m PAH emission. These emissions trace neutral and small PAHs and ionized and larger ones, respectively \citep{Draine21}. 
In Fig.~\ref{fig:PAHinSF_MIRI}, we show the normalized SB profiles at MIRI F770W resolution of the three representative star-forming regions shown in Fig.~\ref{fig:PAHinSF}. The entire procedure used to construct these profiles is analogous to the one presented in Fig.~\ref{fig:PAHinSF}. Because of the different pixel scales, we used a step of 3.5 pixels for the circular annulus apertures which corresponds approximately to the FWHM PSF of F770W. 
Focusing on the SB profiles, we notice that the 3.3 $\mu$m and the 7.7 $\mu$m PAH emission profiles follow each other for 95\% of the star-forming regions. 
We discuss these results in the next section, exploring the scenario of PAH destruction inside star-forming regions.

\section{Discussion} 
\label{sec:discussion}

\subsection{Variations as a function of galactocentric distances}
In this work, we have analyzed HII regions and PAH emission of evolving star-forming regions in NGC 628. We identified them through the Pa$\alpha$ and the 3.3 $\mu$m PAH emission maps. The study is conducted at the highest common physical scales achievable for both the ionized gas tracers and 3.3 $\mu$m one. We have mapped emerging HII regions and their relationship with the associated PDRs down to a resolution of $\sim 7-10$ parsecs. Focusing on the two different tracers (middle panels of Fig.~\ref{fig:rgbFULL}), we notice that the PDRs traced by 3.3 $\mu$m PAH emission are more extended and diffuse than the ionized gas counterpart. As several authors already observed \citep{Boselli04,Bendo08,Crocker13,Calapa14,Pathak23,Sandstrom23b}, less energetic photons are still able to excite PAHs vibrational emission far away from HII regions, making it a strong tracer of diffuse gas in the ISM of local galaxies.


As revealed by the analysis of HII regions and 3.3 $\mu$m PDRs in Sec.~\ref{sec:methods:properties}, their observed properties on average correlate with the galactocentric distance. Brighter and more extended HII regions are found at the location of the co-rotation radius of the spiral pattern. NGC 628 is a pure-spiral arm system with no bar or circumnuclear star-forming ring. The inner regions of the galaxies are dynamically dominated by a stellar bulge which acts as a stabilizer of the gas in the central region of the galaxy \citep{Martig09,Lomaeva22}. In this scenario, the presence of the bulge significantly diminishes the efficiency of the gas to form bound structures, resulting in a noticeable decrease in star formation \citep{Davis22}. This is not the case in the co-rotation radius of the spiral pattern, where gas is piled up, fragments and collapse to form more massive star-forming regions.


\subsection{The gradual morphological evolution of PDRs}
We then focus our analysis on the star-forming regions in order to map the gradual transformation of the natal cloud, where the emerging stellar population has recently formed. Studies conducted with HST \citep[e.g.,][]{Whitmore11,Hannon22} have used the optical appearance of HII regions (compact vs open) as an indicator of temporal evolution. JWST has now opened a new window into the emerging phase, which has remained so far undetected. We focus on the earliest stages, selecting star-forming regions that host a compact HII region and mapping the gradual evolution of the associated PDRs, as traced by PAH emission at unprecedented resolution. We have defined three morphological classes within our sample of star-forming regions: compact, extended, and open PAH profiles. 

The morphological classifications presented in this study unveils various 3.3 $\mu$m PAH distributions in star-forming regions, pointing towards a fast dynamical evolution of the PDRs in the presence of ionizing photons. This is in agreement with previous studies in the Milky Way performed by \cite{Churchwell09} and \cite{Relano09}.
By construction, the ionized gas traced by Pa$\alpha$ and the Br$\alpha$ emission is still abundant and compact in the immediate volume around the newly formed stellar association. 

\begin{figure*}[htb!]
    \centering
    \includegraphics[width = 0.496\textwidth]{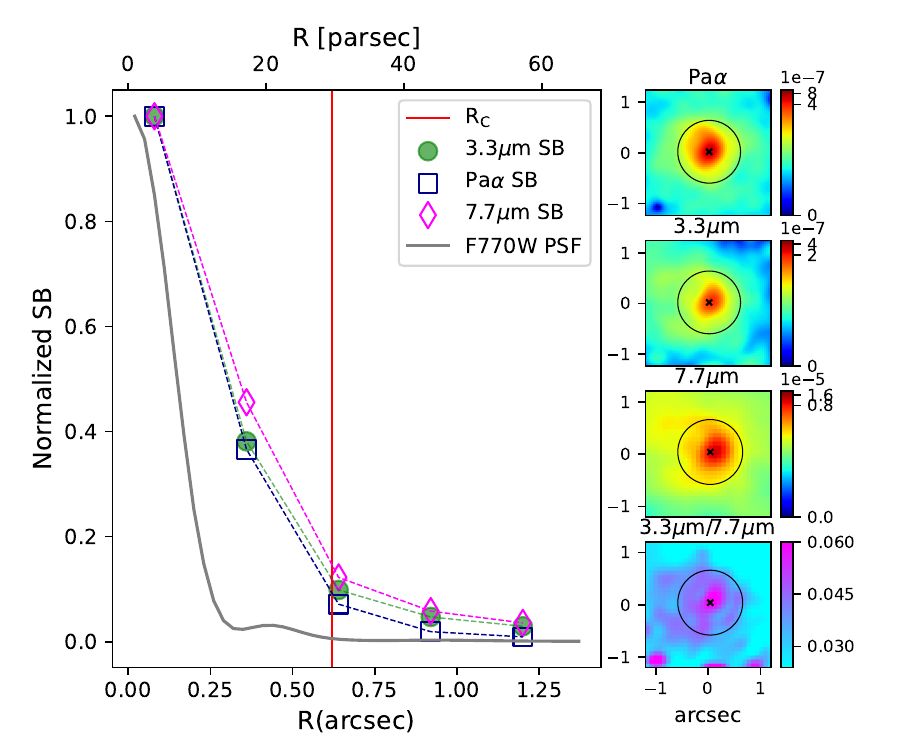}
    \hfill
    \includegraphics[width = 0.496\textwidth]{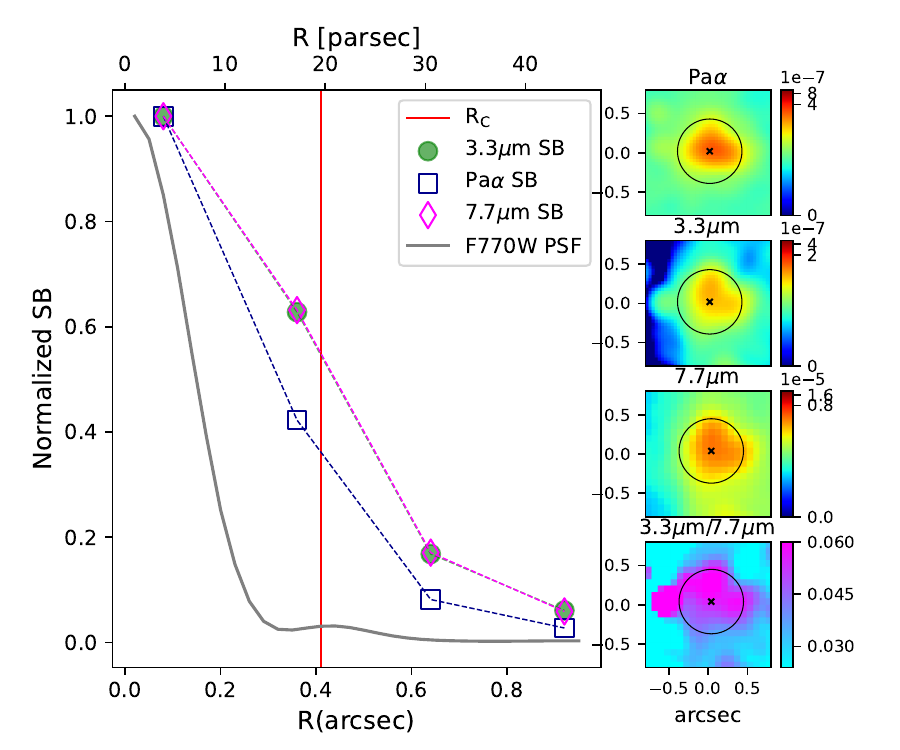}
    \\[\smallskipamount]
\includegraphics[width = 0.496\textwidth]{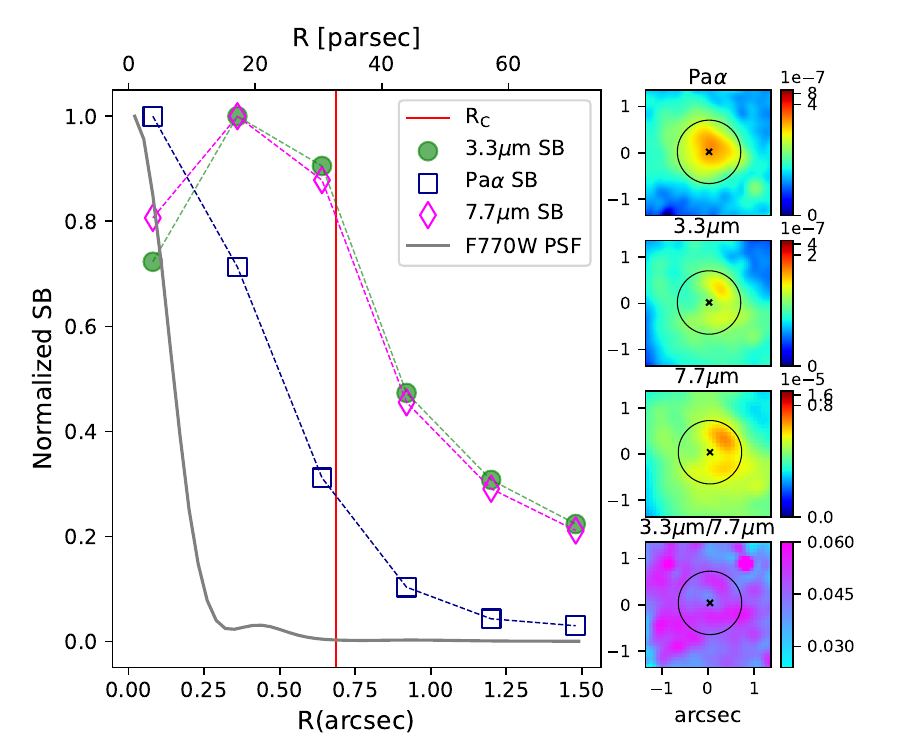}
\caption{Normalized SB profiles of the star-forming regions presented in Fig.~\ref{fig:PAHinSF} at the MIRI F770W resolution, shown in the different sub-panels. For each region, the left panel displays the SB profiles of the Pa$\alpha$ emission (dark blue unfilled squares), the 3.3 $\mu$m PAH emission (green filled circles) and the 7.7 $\mu$m PAH (magenta unfilled diamonds). The grey line represents the normalized SB profile of the PSF in the F770W filter. 
Red vertical solid lines represent the circular radius R$_{\rm C}$ of each region, as described in the text. The four right panels show a zoom-in in each region in the Pa$\alpha$, 3.3 $\mu$m and 7.7 $\mu$m PAH emission maps (log scale, units are {\rm Jy}), as well as the 3.3 $\mu$m/7.7 $\mu$m ratio in the bottom panel. Black crosses are in the center of the region, corresponding to the brightest pixel in the Pa$\alpha$ map. For each panel, the black circle is centered in the center of the region, with a radius of R$_C$.
    }
    \label{fig:PAHinSF_MIRI}
\end{figure*}

In our analysis, we find that the Pa$\alpha$ and Br$\alpha$ normalized SB profiles clearly follow each other, as expected due to the same physical scenario which produces these emission lines in star-forming regions.
On the other hand, the 3.3 $\mu$m SB profiles can be compact at the physical scales resolved in this study ($\sim 7$ pc), following the ionized hydrogen tracers, but can also be deficient in the inner regions. These variations in  morphological appearance are also confirmed when we analyse the 7.7 $\mu$m emission, a tracer for larger and more ionized PAHs \citep{Draine21}, albeit at lower physical resolution ($\sim 12$ pc). 
The similar SB profiles between the two PAH emissions presented in Fig.~\ref{fig:PAHinSF_MIRI} might suggest that in spite of the different physical properties and ionisation state of the molecules responsible for the two bands, they are affected in the same way in proximity of the source of ionizing radiation. This result does not exclude that differences may still arise in the very inner regions, beyond the MIRI resolution.
\\

The observed variations in morphologies point towards an evolution of the PAH distribution around star-forming regions when the ionized gas is still compact and has not expanded yet. In this scenario, PAHs in the proximity of a star-forming region are destroyed in the first 6 Myr of the cluster lifetime. 
We further constrain this by comparing the fitted ages of the ionizing star clusters with the observed PAH emission. 
In Sec.~\ref{sec:res} we showed that older clusters are in general associated with the open morphology distribution, while the compact and extended ones are in general younger. We do not observe a large difference in the mass distributions.

To further test the observed trends, we run Kolmogorov-Smirnov (KS) tests for both the age and the mass distributions, to estimate the probability that the data may originate from the same distribution.
For the mass distributions, we found that the open morphology is likely drawn from a different distribution than the extended class, and even more so for the compact class. The KS (p$_{\rm val}$) for open vs extended and compact classes is 0.13 (0.01) and 0.24 (7.3e-08), respectively. The lower average mass for open morphologies may be explained by either the smaller sample size or the fact that more massive cluster potentially evolve faster and do not enter anymore into our classification, since the gas has been cleared and the ionized emission is not observed as compact anymore. This sample is by construction biased against star-forming regions where the ionized gas is in a shell. The dependency of this evolution on the cluster mass will be further explored in upcoming studies where the NGC 628 sample will be combined with those obtained from the other two FEAST spiral galaxies, M83 and M51, which will substantially increase the statistic. The mass range, and possibly environment and metallicity, will be different and this may allow us to tie down the role of cluster mass in determining the timescale for cluster emergence.

Regarding the age of the associated eYSCs, while the extended and the compact distributions are likely to be drawn from the same distribution (KS = 0.08, p$_{\rm val}$ = 0.20), the open one lies in a different space, with KS values of 0.15 and 0.20, and p$_{\rm val}$ of 0.004 and 1.1e-05 for tests against extended and compact distributions.
The eYSCs associated with an open morphology are statistically older than the compact counterpart, possibly suggesting that the morphological evolution of PDRs is linked to the feedback processes by young and massive stars, which lead to a projected spatial decoupling of PAH grains and HII regions after $\sim$4 Myr. 
Geometrical effects, mentioned in Sec.~\ref{sec:SFandPDR}, are unavoidable and naturally linked to the star formation process. We expect them to be averaged out, as we provide a statistical sample. 
Finally, the observed drop in all the distributions at 6 Myr is a physical phenomenon. After approximately 6-7 Myr, identifying sources that emit recombination lines is challenging because the clusters are sufficiently old to have a significantly reduced ionizing photon flux ($\sim$1/100th), compared to clusters at 2-3 Myr \citep{Calzetti10}. 

The distribution of the nebular E(B-V) values reveals that we do not find correlations between attenuation and the three PAH profiles. These findings possibly evidence that, in the assumption of a dust screen geometry, the stochastic dust emission powered by PAHs and the relative evolution of PDRs do not play a key role in obscuring the stellar light during the first Myrs. The attenuation is indeed dominated by the thermal dust emission. \cite{Povich07,Relano09}, among many others, showed that in star-forming regions where the PAH emission is in a shell, hot dust is still co-spatial with the ionized gas, providing the attenuation. This issue will be further investigated in Linden et al.~(in prep.). 

\subsection{PAH destruction}
In our study, the analysis of the PAH morphologies is consistent with an evolutionary sequence where PAHs are destroyed by stellar feedback at physical scales of a few parsecs. 
While the HII regions are still compact and the stellar feedback has not yet dissolved them, open PAH morphologies are associated with older eYSCs which have been destroying PAH molecules over time.
In this analysis, the extended class may represent an intermediate stage that is fully consistent with the evolutionary sequence we defined. Our results are consistent with \cite{Relano09}, who noticed the lack of PAH in HII regions traced by H$\alpha$ emission.
\\
\begin{figure}[htb]
    \centering
    \includegraphics[width = \columnwidth]{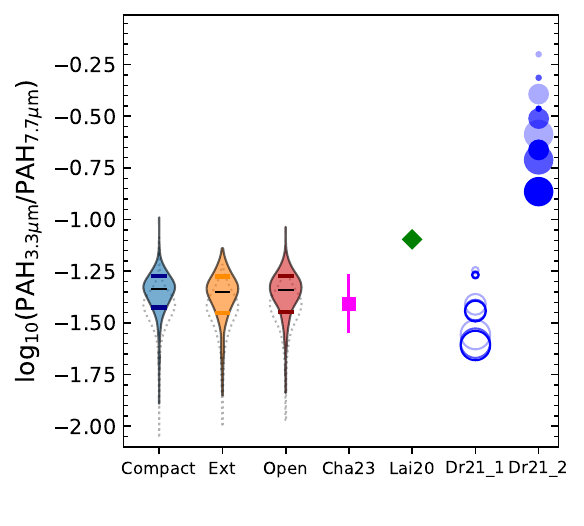}
    \caption{Observations at different physical scales and emission estimated from models of the PAH$_{3.3 \mu{\rm m}}$/PAH$_{7.7 \mu{\rm m}}$ ratio. 
    From left to right: PAH$_{3.3 \mu{\rm m}}$/PAH$_{7.7 \mu{\rm m}}$ ratio distributions of our sample of star-forming regions, color-coded as in Fig.~\ref{fig:massAge_CIGALE}. The flux estimation is described in the text. For each violin, the horizontal black line represents the median value of the distribution, while the colored ones represent the 16th and 84th percentiles. Beneath them, the dotted violins represent the same distribution without the subtraction of the continuum to the 7.7 $\mu$m PAH emission map using the F560W filter.
    The magenta square represents the median value of this ratio founded by \cite{Chastenet23B} for the total emission in NGC 628. Errorbars represent the 16th-84th percentile of the distribution.
    The green diamond represents the ratio for the star-forming 1C galaxy sample in \cite{Lai20}, where the spectrum has been convolved to the JWST filter throughputs and the stellar continuum has been removed from it. Errorbars lie within the symbol.
    Blue circles show PAH band ratios for the models of \cite{Draine21} convolved to the JWST filter throughputs, using a 3 Myr-old starburst as illuminating starlight spectrum and $U = 1$. A different size and opacity of a circle represents a different grain size distribution or a different ionization model: bigger circles correspond to bigger grains and higher opacity corresponds to higher ionization. The unfilled circles (Dr21\_1) correspond to models considering ionized PAHs and astrodust, while the filled ones (Dr21\_2) represent models considering also neutral PAHs.}
    \label{fig:qPAH_CIGALE_draine}
\end{figure}

To further test whether PAH disruption is the mechanism responsible for the morphological evolution of star-forming regions and to investigate the relationship between different species of PAHs, we plot the measured PAH$_{3.3 \mu{\rm m}}$/PAH$_{7.7 \mu{\rm m}}$ ratio in the 3 classes in Fig.~\ref{fig:qPAH_CIGALE_draine}, including data and models from the literature.

\begin{figure*}[htb]
    \centering
    \includegraphics[width = 1.0\textwidth]{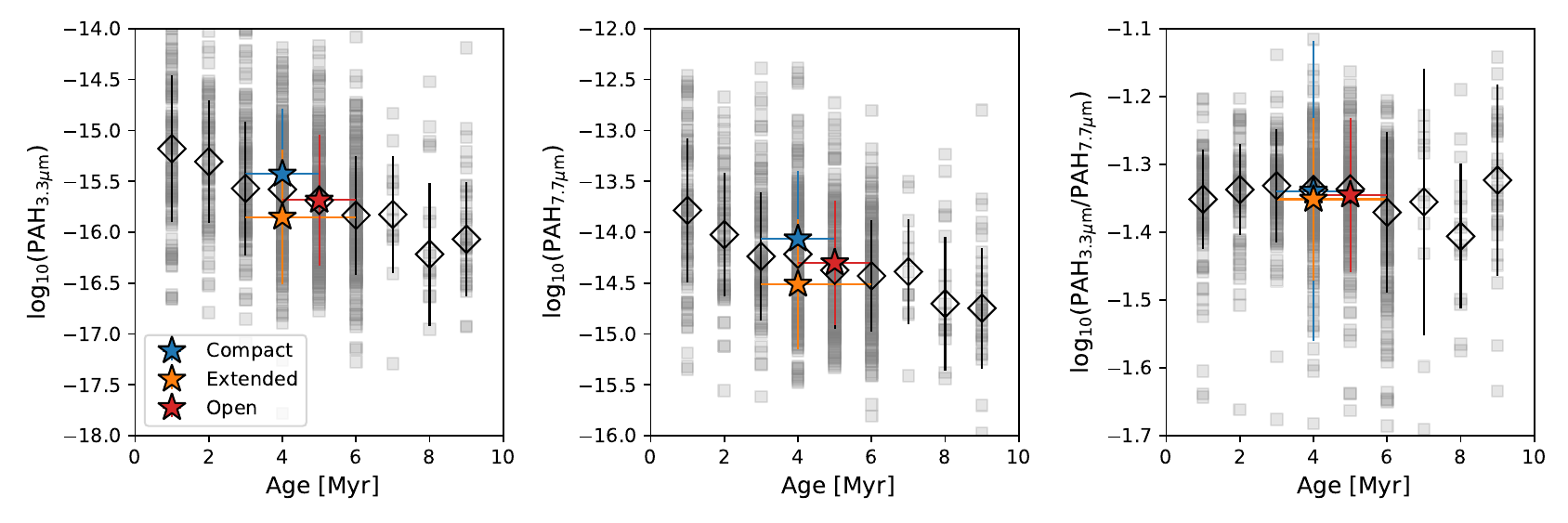}
    \caption{PAH$_{3.3 \mu{\rm m}}$ flux ({\it left panel}), PAH$_{7.7 \mu{\rm m}}$ flux ({\it middle panel}) and PAH$_{3.3 \mu{\rm m}}$/PAH$_{7.7 \mu{\rm m}}$ flux ratio ({\it right panel}) as a function of the eYSCs derived ages for our sample of star-forming regions. Flux units are erg s$^{-1}$ cm$^{-2}$. The flux estimation using aperture photometry is described in the text. Black unfilled diamonds represent median values of the fluxes in different bins of ages of 1 Myr. Filled stars represent median values  of the same distribution binned in according with our morphological classification and color-coded as in Fig.~\ref{fig:massAge_CIGALE}, where errorbars on the age axes correspond to the 25\% and 75\% percentiles, as shown by the boxplots in Fig.~\ref{fig:massAge_CIGALE}.
    Errorbars on the flux axes indicate the standard deviation of each binned distribution.}
    \label{fig:ratio_age_atrad}
\end{figure*}

We measure the PAH$_{3.3 \mu{\rm m}}$/PAH$_{7.7 \mu{\rm m}}$ ratio for each star-forming region using aperture photometry, where the radius of the circular aperture is the aforementioned ${\rm R}_{\rm C}$. We observe no significant differences in the ratio distribution when using the non-subtracted 7.7 $\mu$m map (represented by the dotted empty violins in Fig.~\ref{fig:qPAH_CIGALE_draine}). This suggests that the observed ratio is primarily influenced by the two PAH features.
In the figure, we show the recovered values color-coded by the different morphologies as in Fig.~\ref{fig:morphoClass}. We include values of the sample of star-formation dominated galaxies with prominent PAH emission in \cite{Lai20}, as well as the median value for this ratio that \cite{Chastenet23B} found for NGC 628, when considering the total contribution from HII and diffuse regions. Finally, we add PAH ratios from \cite{Draine21} models, for various assumptions of size distributions and ionization parameters. These models are derived for a diffuse ISM and the resulting PAH spectra do not include stellar emission.
We show models considering emission from both neutral and ionized PAHs (Dr21\_2 in Fig.~\ref{fig:qPAH_CIGALE_draine}) and models taking into account only emission from ionized PAHs (Dr21\_1 in Fig.~\ref{fig:qPAH_CIGALE_draine}).
We observe that PAH ratios are generally constant among the three morphology classifications, evidencing that the molecules emitting at 3.3 and 7.7 $\mu$m may undergo the same destruction process, as already suggested by the results on the SB profiles presented in Fig.~\ref{fig:PAHinSF_MIRI}. In Fig.~\ref{fig:ratio_age_atrad}, we present the 3.3 $\mu$m and 7.7 $\mu$m fluxes as a function of the age of the eYSCs for our sample of star-forming regions. We observe a consistent trend where the 3.3 $\mu$m and 7.7 $\mu$m fluxes exhibit an anti-correlation with the age of the eYSCs, resulting in constant ratios when the sample is divided by morphology. Furthermore, when compared to PAH models from \cite{Draine21}, we see that these predict higher PAH ratios if we consider both neutral and ionized PAH emission (Dr21\_2 in Fig.~\ref{fig:qPAH_CIGALE_draine}), while the models that include only ionised PAH are consistent with the values we observe. The recovered ratios are similar to those derived by \citep{Chastenet23B} at comparable physical scales and lower than reported by \citet{Lai20} at galactic scales. 
The 7.7 $\mu$m PAH emission traces more ionized and colder PAHs \citep{Draine21}, which implies that the emission in the latter PAH is enhanced with respect the 3.3 $\mu$m band, in agreement with the trends in the models where only ionised PAH are included (Dr21\_1 in Fig.~\ref{fig:qPAH_CIGALE_draine}). 
We might also be seeing an effect of grain processing, where smaller PAH molecules traced by 3.3 $\mu$m emission are preferentially destroyed closer to the HII regions \citep{Leger89b}. This consideration holds in the case where the same mechanism is responsible of destroying PAHs which emit at 3.3 and 7.7 $\mu$m during the temporal evolution of the star-forming region. Furthermore, more excitation effects can play a role in the interpretation of the 3.3 $\mu$m emission, as discussed in \cite{Draine21}. 

\section{summary}
\label{sec:end}
In this paper, we have performed a morphological analysis at a few parsecs resolution of a sample of star-forming regions using JWST NIRCam observations of Pa$\alpha$, Br$\alpha$ and 3.3 $\mu$m PAH, in addition to the MIRI 7.7 $\mu$m PAH. Thanks to the unprecedented resolution provided by JWST, we have mapped PDR evolution at scales of $\sim10$ pc. 
From a sample of compact HII regions, we have characterised three different morphologies of PDRs traced by 3.3 $\mu$m PAH emission. We found that:

\begin{itemize}
    \item The temporal evolution of the PDRs, traced by the associated eYSCs, within star-forming regions, correlates with the distribution and morphology of the PAH emission. Regions with a less compact morphology are statistically older than their compact counterpart. In particular, open regions span an age range between 3 and 6 Myr with a median value of 5 Myr, while the compact ones present a broader range between 1 and 6 Myr, with the median value being 4 Myr. 

    \item 
    We found lower values of the PAH$_{3.3 \mu{\rm m}}$/PAH$_{7.7 \mu{\rm m}}$ ratio when compared to PAH models \citep{Draine21} and integrated emission from galaxies. The observed difference might be explained by both the presence of more ionized PAHs and the effects of grain processing in the destruction mechanisms.
    \item While the 3.3 $\mu$m and 7.7 $\mu$m emissions decrease as a function of cluster age, their ratio is constant among the three classes. SB profiles of the two PAH track each other suggesting that PAHs emitting at 3.3 $\mu$m (smaller and neutral) and 7.7 $\mu$m (larger and ionized) may be undergoing the same general destruction mechanism.
    \item We did not find a correlation between the morphological classes and the eYSCs attenuation, possibly suggesting that during this emerging phase, the attenuation of the stellar light does not strongly depend on the evolutionary stage of the PDR but more on geometrical effects.

\end{itemize}

This work has probed for the first time the temporal evolution of PDRs, traced by PAH emission, at sub-10 pc scales outside of the Local Volume. Follow-up analyses and observations of the other FEAST galaxies will shed light to the complex mechanism of PAH destruction in proximity of newly formed star clusters in different galactic environments and metallicities. 

\section*{acknowledgments}
The authors would like to express their gratitude to the anonymous referee for their insightful comments, which have significantly improved the content of this paper.
This work is based in part on observations made with the NASA/ESA/CSA James Webb Space Telescope, which is operated by the Association of Universities for Research in Astronomy, Inc., under NASA contract NAS 5-03127. The data were obtained from the Mikulski Archive for Space Telescopes (MAST) at the Space Telescope Science Institute.
The observations presented in this work are associated with program \# 1783. Support for this program was provided by NASA through a grant from the Space Telescope Science Institute, which is operated by the Association of Universities for Research in Astronomy, Inc., under NASA contract NAS 5-03127. The specific observations analyzed can be accessed via\dataset[10.17909/3xyt-7h62]{https://doi.org/10.17909/3xyt-7h62}. 
A.A and A.P. acknowledge support from the Swedish National Space Agency (SNSA) through the grant 2021- 00108.
A.P. would like to express gratitude to Thomas Lai for his feedback on this work.
KG is supported by the Australian Research Council through the Discovery Early Career Researcher Award (DECRA) Fellowship (project number DE220100766) funded by the Australian Government. 
KG is supported by the Australian Research Council Centre of Excellence for All Sky Astrophysics in 3 Dimensions (ASTRO~3D), through project number CE170100013.
This research made use of astrodendro, a Python package to compute dendrograms of Astronomical data (http://www.dendrograms.org/)
Models from \cite{Draine21} are publicly available at: {\url{https://dataverse.harvard.edu/dataset.xhtml?persistentId=doi:10.7910/DVN/LPUHIQ}} \citep{DraineModels}.

\appendix
\section{2D normalization of the continuum subtracted maps}
\label{bkgNorm}

\begin{figure*}[htb!]
    \centering
    \includegraphics[width = \textwidth]{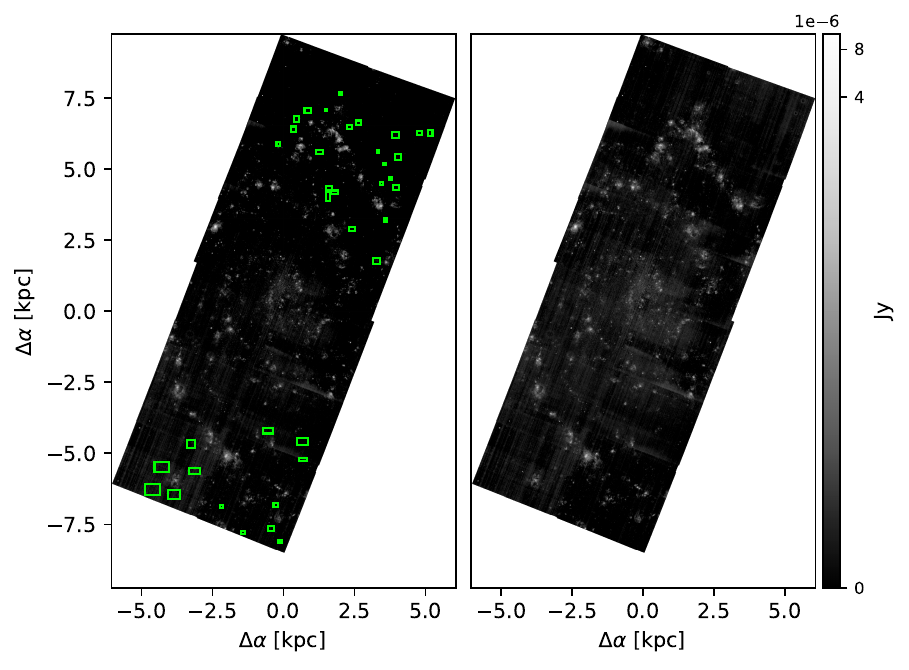}

    \caption{{\it Left}: Pa$\alpha$ emission line map of NGC 628 before the image has been normalized to 0 by the fitting method described in the text. Green boxes represent the empty region of the sky utilized to estimate the background image. {\it Right}: Resulting Pa$\alpha$ map after subtraction of the modeled background image.} 
    
    \label{fig:bkg187}
\end{figure*}
\begin{figure*}[htb!]
    \centering
    \includegraphics[width = \textwidth]{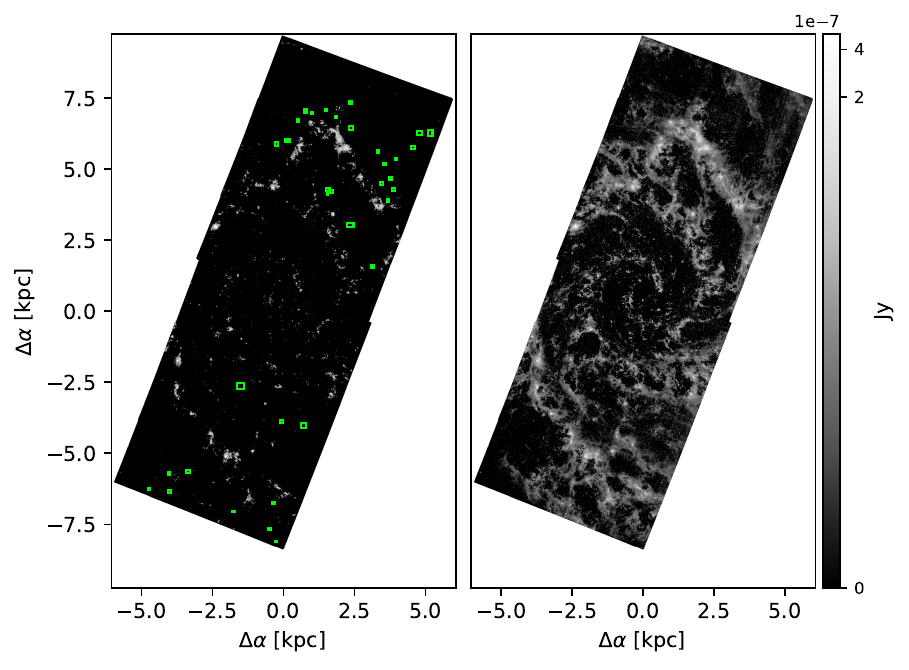}

    \caption{{\it Left}: 3.3 $\mu$m emission map of NGC 628 before the image has been normalized to 0 by the fitting method described in the text. Green boxes represent the empty region of the sky utilized to estimate the background image. {\it Right}: Resulting 3.3 $\mu$m map after subtraction of the modeled background image.} 
    
    \label{fig:bkg335}
\end{figure*}

As described in the text, we corrected a residual background gradient in the continuum subtracted NIRCam maps by subtracting a 2D plane. The 2D plane is created by manually selecting a set of empty sky regions. We show the empty regions in the left panels of Fig.~\ref{fig:bkg187} and Fig.~\ref{fig:bkg335} for the Pa$\alpha$ and 3.3 $\mu$m emission maps, respectively. 
For each region, we estimated the mode and RMS values by fitting a Gaussian profile to the count distribution within the region. 
Afterward, we fitted a 2D plane using the mode values of the regions to obtain a model of the background image.
We then subtracted this fitted background image, removing the residual background gradient. After this residual background correction, the mode of the background regions is consistent with zero. The resulting emission line maps presented in the right panels of Fig.~\ref{fig:bkg187} and Fig.~\ref{fig:bkg335}. 

\begin{figure*}[htb!]
    \centering
    \includegraphics[width = \textwidth]{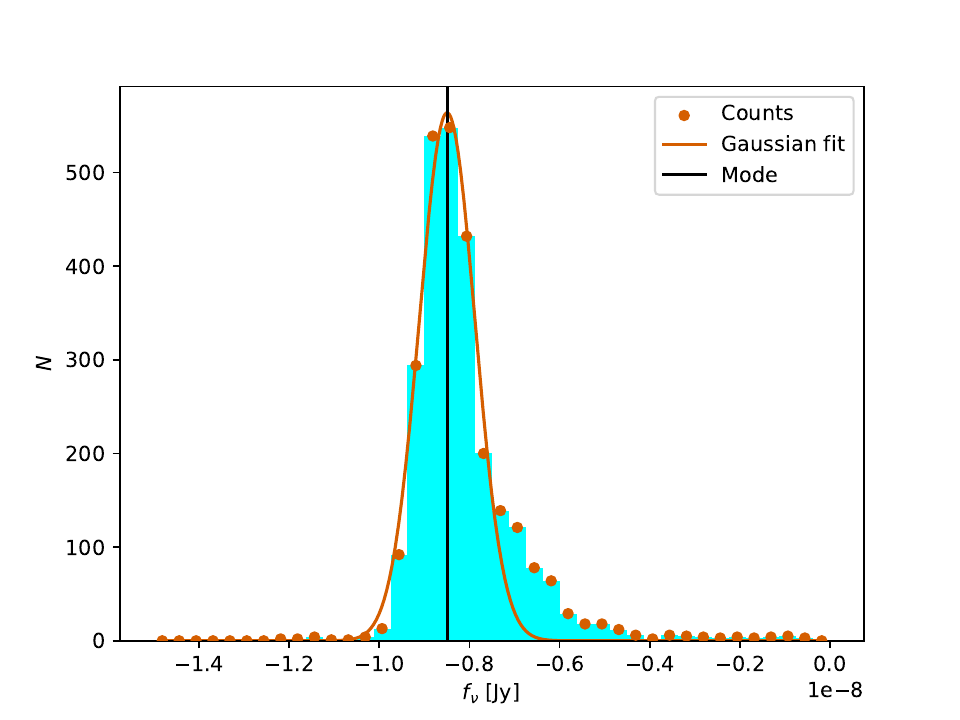}

    \caption{Count distribution in an representative empty region of the sky in the 3.3 $\mu$m emission map, showing contamination due to the diffuse nature of the PAH emission, seen as the tail in the flux distribution towards high flux density. Note the Gaussian fit provides a reasonable estimation of the background level, even when diffuse PAH emission contaminates the regions.} 
    
    \label{fig:Gaussfit_ex}
\end{figure*}

The diffuse nature of the 3.3 $\mu$m PAH emission feature might have caused flux contamination in the selected sky regions, leading to an over-estimation of the RMS. However, we note that fitting a Gaussian profile can efficiently avoid this issue in regions where we observe flux contamination in the 3.3 $\mu$m emission, as shown in the example of Fig.~\ref{fig:Gaussfit_ex}.

\section{Attenuation Issue}
\label{attenuation}

\begin{figure*}[htb!]
    \centering
    \includegraphics[width = 0.496\textwidth]{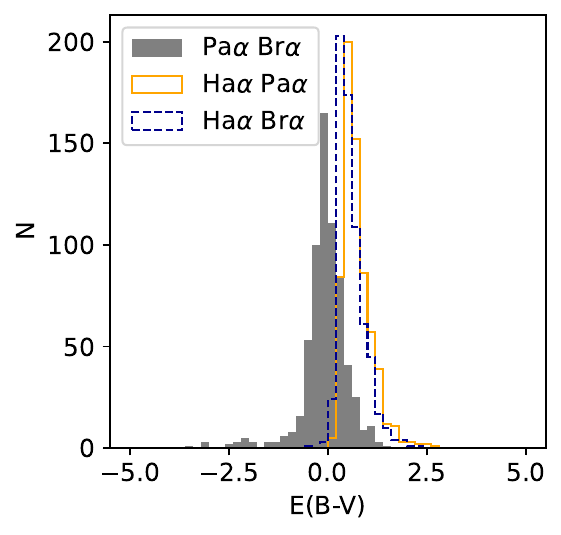}
    \hfill
    \includegraphics[width = 0.496\textwidth, trim={0.41cm 0cm 0cm 0cm}]{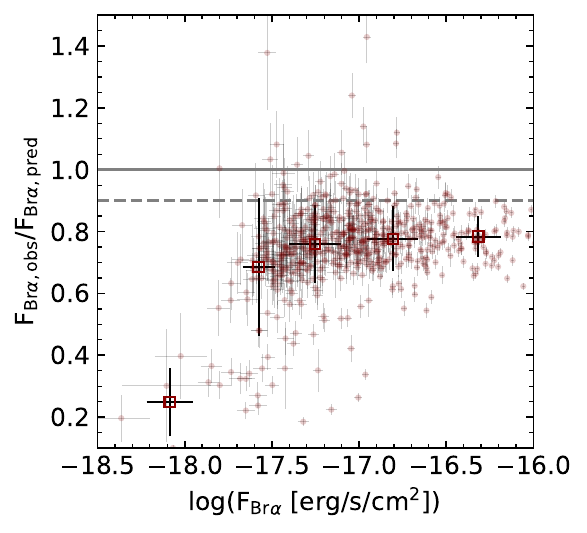}

    \caption{{\it Left}: E(B-V) distribution for our sample of HII regions located within the HST and JWST field of view. The three different histograms represent attenuation values estimated using the flux ratio between Pa$\alpha$/Br$\alpha$, H$\alpha$/Pa$\alpha$ and H$\alpha$/Br$\alpha$. {\it Right}: observed Br$\alpha$ flux to predicted Br$\alpha$ flux ratio vs observed Br$\alpha$ flux. The grey solid line corresponds to a constant ratio of 1, while the dashed one corresponds to a constant ratio of 0.9. The red unfilled squares represent the median values in bins of observed $\log( {\rm F}_{{\rm Br}\alpha})$ .We provide a description of the estimation of the predicted Br$\alpha$ flux in the text.} 
    
    \label{fig:ebv_ha}
\end{figure*}

As mentioned in the text, the ratio Pa$\alpha$ to Br$\alpha$ reveals a significant number of values with low or even negative values of attenuation. We investigated this issue measuring E(B-V) values for our sample of HII regions. 
To measure the flux ratio between the two emission features, we employed a circular aperture of 3 pixels (0.12$\arcsec$ $\sim$ 5.72 pc) in radius to perform aperture photometry. We also corrected for the local background in the same way as described in the text. We then estimated the relative attenuation using the \cite{Fahrion23} extinction law for 30 Doradus and assuming case B recombination. We assume $n_e = 100$ cm$^{-3}$ and $T = 10000$ K.
Moreover, we used HST ACS F555W, F814W and F658N observations (see Sec.~\ref{sec:obs&dr}) to obtain a continuum subtracted image of the H$\alpha$ emission feature in NGC 628 \citep{Gregg24}, where the science field of view covers most of the JWST field of view. We used the F555W and F814W filters to remove the continuum from the F658N. To obtain a reliable estimation of the H$\alpha$ flux, we also removed the NII contamination from the H$\alpha$ emission line map. Within this map, we estimated the attenuation in our sample of regions using the color excess between H$\alpha$ and the two NIRCam nebular lines.
The grey histogram in the left panel of Fig.~\ref{fig:ebv_ha} shows the distribution of the attenuation values (located within the common field of view of HST and JWST) obtained from the color excess between Pa$\alpha$ and Br$\alpha$. 
The distribution of E(B-V) values is centered near zero, with many regions showing negative values. However, the E(B-V) estimated by means of the H$\alpha$ line are consistent with positive values (unfilled orange and blue histograms), suggesting issues in the flux values of either the Pa$\alpha$ or the Br$\alpha$ emission feature. 
In the right panel of Fig.~\ref{fig:ebv_ha}, we investigated this possible flux issue by measuring the ratio between the observed Br$\alpha$ and the predicted one. The predicted Br$\alpha$ flux is defined under the assumption that the E(B-V) value estimated from the Pa$\alpha$ and Br$\alpha$ lines equals that estimated from the H$\alpha$ and Pa$\alpha$ lines. The plot clearly shows a discrepancy between the two values, with the observed Br$\alpha$ flux being underestimated (or the observed Pa$\alpha$ flux being overestimated) by about 20\%.
The observed discrepancy may arise from various factors, such as flux calibrations, data reduction and subtraction of the continuum, especially in the redder filters. Moreover, the generation of the final mosaic from single detector frames could also affect the flux values.

To exclude the latter, we performed the same analysis comparing the attenuation values obtained for a single detector and the entire mosaic (considering HII regions located within the field of view of the single detector). 
\begin{figure*}[htb!]
    \centering
    \includegraphics[width = 0.8\textwidth]{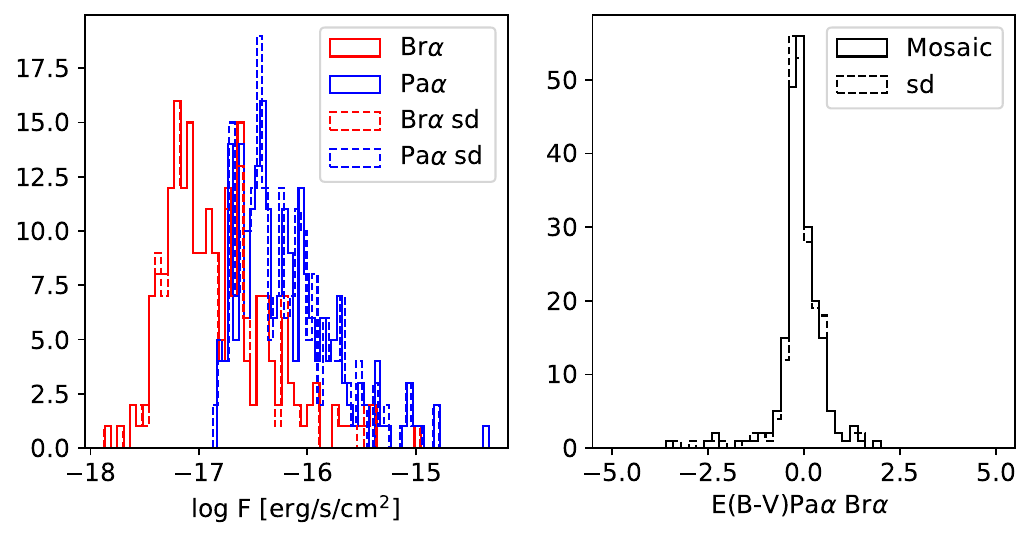}\\[\smallskipamount]
    
    \includegraphics[width =0.8\textwidth]{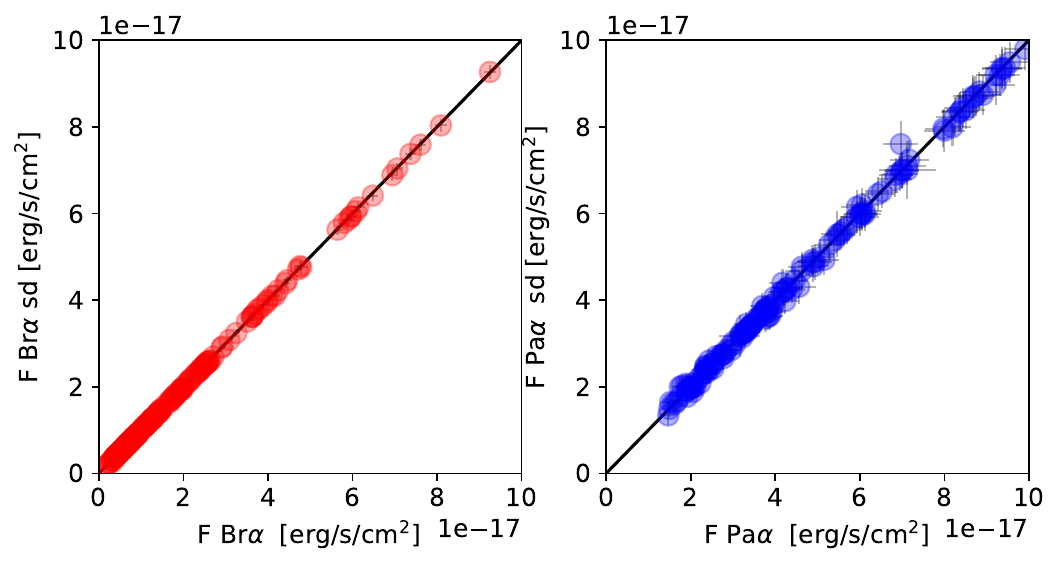}
    
\caption{{\it Top}: Pa$\alpha$ and Br$\alpha$ fluxes and respective E(B-V) distributions from Pa$\alpha/$Br$\alpha$  for both the entire mosaic and the single detector (identified in the legend with the keyword ``sd'') in our sample of HII regions located within the single detector field of view.
{\it Bottom}: Single detector and entire mosaic flux comparison for Pa$\alpha$ and Br$\alpha$. The solid lines represent the one to one relation.
    }
    \label{fig:ebv_sd}
\end{figure*}
We show the results of this analysis in Fig.~\ref{fig:ebv_sd}. In the top panels, we present the distribution of the Pa$\alpha$ and Br$\alpha$ fluxes and of the E(B-V) for the single detector and the entire mosaic, while in the bottom panels we show flux comparisons. 
These plots clearly reveal that we do not witness a significant difference between the fluxes in the single frame and in the entire mosaic. For this reason, we can exclude the contribution of the mosaic assembly to the unphysical attenuation values.

In conclusion, our analysis may point toward an issue in the estimation of the stellar continuum at longer wavelength, where it can be contaminated by the presence of hot dust emission. We remain abreast of the latest developments in the JWST pipeline release and follow-up works will give us more information about this issue.

%


\bibliography{Mypapers}{}

\begin{thebibliography}{}
\expandafter\ifx\csname natexlab\endcsname\relax\def\natexlab#1{#1}\fi
\providecommand{\url}[1]{\href{#1}{#1}}
\providecommand{\dodoi}[1]{doi:~\href{http://doi.org/#1}{\nolinkurl{#1}}}
\providecommand{\doeprint}[1]{\href{http://ascl.net/#1}{\nolinkurl{http://ascl.net/#1}}}
\providecommand{\doarXiv}[1]{\href{https://arxiv.org/abs/#1}{\nolinkurl{https://arxiv.org/abs/#1}}}

\bibitem[{{Allain} {et~al.}(1996){Allain}, {Leach}, \& {Sedlmayr}}]{Allain96}
{Allain}, T., {Leach}, S., \& {Sedlmayr}, E. 1996, \aap, 305, 602

\bibitem[{Bajaj(2023)}]{VarunSoftware}
Bajaj, V. 2023, {JWST Mosaic Sky Match}, 0.1.
\newblock \url{https://github.com/Vb2341/jwst-mosaic-skymatch}

\bibitem[{{Bendo} {et~al.}(2008){Bendo}, {Draine}, {Engelbracht}, {Helou}, {Thornley}, {Bot}, {Buckalew}, {Calzetti}, {Dale}, {Hollenbach}, {Li}, \& {Moustakas}}]{Bendo08}
{Bendo}, G.~J., {Draine}, B.~T., {Engelbracht}, C.~W., {et~al.} 2008, \mnras, 389, 629, \dodoi{10.1111/j.1365-2966.2008.13567.x}

\bibitem[{{Boquien} {et~al.}(2019){Boquien}, {Burgarella}, {Roehlly}, {Buat}, {Ciesla}, {Corre}, {Inoue}, \& {Salas}}]{Boquien19}
{Boquien}, M., {Burgarella}, D., {Roehlly}, Y., {et~al.} 2019, \aap, 622, A103, \dodoi{10.1051/0004-6361/201834156}

\bibitem[{{Boselli} {et~al.}(2004){Boselli}, {Lequeux}, \& {Gavazzi}}]{Boselli04}
{Boselli}, A., {Lequeux}, J., \& {Gavazzi}, G. 2004, \aap, 428, 409, \dodoi{10.1051/0004-6361:20041316}

\bibitem[{{Bradley} {et~al.}(2023){Bradley}, {Sip{\H{o}}cz}, {Robitaille}, {Tollerud}, {Vin{\'\i}cius}, {Deil}, {Barbary}, {Wilson}, {Busko}, {Donath}, {G{\"u}nther}, {Cara}, {Lim}, {Me{\ss}linger}, {Conseil}, {Bostroem}, {Droettboom}, {Bray}, {Andersen Bratholm}, {Jamieson}, {Ginsburg}, {Barentsen}, {Craig}, {Morris}, {Perrin}, {Rathi}, {Pascual}, {Perren}, {Georgiev}, \& {Kerzendorf}}]{Bradley23}
{Bradley}, L., {Sip{\H{o}}cz}, B., {Robitaille}, T., {et~al.} 2023, {astropy/photutils: 1.9.0}, 1.9.0, Zenodo,  Zenodo, \dodoi{10.5281/zenodo.596036}

\bibitem[{{Calapa} {et~al.}(2014){Calapa}, {Calzetti}, {Draine}, {Boquien}, {Kramer}, {Xilouris}, {Verley}, {Braine}, {Rela{\~n}o}, {van der Werf}, {Israel}, {Hermelo}, \& {Albrecht}}]{Calapa14}
{Calapa}, M.~D., {Calzetti}, D., {Draine}, B.~T., {et~al.} 2014, \apj, 784, 130, \dodoi{10.1088/0004-637X/784/2/130}

\bibitem[{{Calzetti}(2013)}]{Calzetti13}
{Calzetti}, D. 2013, in Secular Evolution of Galaxies, ed. J.~{Falc{\'o}n-Barroso} \& J.~H. {Knapen}, 419, \dodoi{10.48550/arXiv.1208.2997}

\bibitem[{{Calzetti} {et~al.}(2000){Calzetti}, {Armus}, {Bohlin}, {Kinney}, {Koornneef}, \& {Storchi-Bergmann}}]{Calzetti00}
{Calzetti}, D., {Armus}, L., {Bohlin}, R.~C., {et~al.} 2000, \apj, 533, 682, \dodoi{10.1086/308692}

\bibitem[{{Calzetti} {et~al.}(2010){Calzetti}, {Chandar}, {Lee}, {Elmegreen}, {Kennicutt}, \& {Whitmore}}]{Calzetti10}
{Calzetti}, D., {Chandar}, R., {Lee}, J.~C., {et~al.} 2010, \apjl, 719, L158, \dodoi{10.1088/2041-8205/719/2/L158}

\bibitem[{{Calzetti} {et~al.}(2015){Calzetti}, {Lee}, {Sabbi}, {Adamo}, {Smith}, {Andrews}, {Ubeda}, {Bright}, {Thilker}, {Aloisi}, {Brown}, {Chandar}, {Christian}, {Cignoni}, {Clayton}, {da Silva}, {de Mink}, {Dobbs}, {Elmegreen}, {Elmegreen}, {Evans}, {Fumagalli}, {Gallagher}, {Gouliermis}, {Grebel}, {Herrero}, {Hunter}, {Johnson}, {Kennicutt}, {Kim}, {Krumholz}, {Lennon}, {Levay}, {Martin}, {Nair}, {Nota}, {{\"O}stlin}, {Pellerin}, {Prieto}, {Regan}, {Ryon}, {Schaerer}, {Schiminovich}, {Tosi}, {Van Dyk}, {Walterbos}, {Whitmore}, \& {Wofford}}]{Calzetti15}
{Calzetti}, D., {Lee}, J.~C., {Sabbi}, E., {et~al.} 2015, \aj, 149, 51, \dodoi{10.1088/0004-6256/149/2/51}

\bibitem[{{Calzetti} {et~al.}(2023){Calzetti}, {Linden}, {McQuaid}, {Messa}, {Ji}, {Krumholz}, {Adamo}, {Elmegreen}, {Grasha}, {Johnson}, {Sabbi}, {Smith}, \& {Bajaj}}]{Calzetti23}
{Calzetti}, D., {Linden}, S.~T., {McQuaid}, T., {et~al.} 2023, \apj, 946, 1, \dodoi{10.3847/1538-4357/acbeac}

\bibitem[{{Camargo} {et~al.}(2015){Camargo}, {Bonatto}, \& {Bica}}]{Camargo15}
{Camargo}, D., {Bonatto}, C., \& {Bica}, E. 2015, \mnras, 450, 4150, \dodoi{10.1093/mnras/stv840}

\bibitem[{{Cepa} \& {Beckman}(1990)}]{Cepa90}
{Cepa}, J., \& {Beckman}, J.~E. 1990, \apj, 349, 497, \dodoi{10.1086/168335}

\bibitem[{{Chastenet} {et~al.}(2023){Chastenet}, {Sutter}, {Sandstrom}, {Belfiore}, {Egorov}, {Larson}, {Leroy}, {Liu}, {Rosolowsky}, {Thilker}, {Watkins}, {Williams}, {Barnes}, {Bigiel}, {Boquien}, {Chevance}, {Dale}, {Kruijssen}, {Emsellem}, {Grasha}, {Groves}, {Hassani}, {Hughes}, {Kreckel}, {Meidt}, {Pan}, {Querejeta}, {Schinnerer}, \& {Whitcomb}}]{Chastenet23B}
{Chastenet}, J., {Sutter}, J., {Sandstrom}, K., {et~al.} 2023, \apjl, 944, L12, \dodoi{10.3847/2041-8213/acac94}

\bibitem[{{Chown} {et~al.}(2023){Chown}, {Sidhu}, {Peeters}, {Tielens}, {Cami}, {Bern{\'e}}, {Habart}, {Alarc{\'o}n}, {Canin}, {Schroetter}, {Trahin}, {Van De Putte}, {Abergel}, {Bergin}, {Bernard-Salas}, {Boersma}, {Bron}, {Cuadrado}, {Dartois}, {Dicken}, {El-Yajouri}, {Fuente}, {Goicoechea}, {Gordon}, {Issa}, {Joblin}, {Kannavou}, {Khan}, {Lacinbala}, {Languignon}, {Le Gal}, {Maragkoudakis}, {Meshaka}, {Okada}, {Onaka}, {Pasquini}, {Pound}, {Robberto}, {R{\"o}llig}, {Schefter}, {Schirmer}, {Vicente}, {Wolfire}, {Zannese}, {Aleman}, {Allamandola}, {Auchettl}, {Baratta}, {Bejaoui}, {Bera}, {Black}, {Boulanger}, {Bouwman}, {Brandl}, {Brechignac}, {Br{\"u}nken}, {Buragohain}, {Burkhardt}, {Candian}, {Cazaux}, {Cernicharo}, {Chabot}, {Chakraborty}, {Champion}, {Colgan}, {Cooke}, {Coutens}, {Cox}, {Demyk}, {Donovan Meyer}, {Foschino}, {Garc{\'\i}a-Lario}, {Gavilan}, {Gerin}, {Gottlieb}, {Guillard}, {Gusdorf}, {Hartigan}, {He}, {Herbst}, {Hornekaer}, {J{\"a}ger}, {Janot-Pacheco}, {Kaufman}, {Kemper}, {Kendrew},
  {Kirsanova}, {Klaassen}, {Kwok}, {Labiano}, {Lai}, {Lee}, {Lefloch}, {Le Petit}, {Li}, {Linz}, {Mackie}, {Madden}, {Mascetti}, {McGuire}, {Merino}, {Micelotta}, {Misselt}, {Morse}, {Mulas}, {Neelamkodan}, {Ohsawa}, {Omont}, {Paladini}, {Palumbo}, {Pathak}, {Pendleton}, {Petrignani}, {Pino}, {Puga}, {Rangwala}, {Rapacioli}, {Ricca}, {Roman-Duval}, {Roser}, {Roueff}, {Rouill{\'e}}, {Salama}, {Sales}, {Sandstrom}, {Sarre}, {Sciamma-O'Brien}, {Sellgren}, {Shenoy}, {Teyssier}, {Thomas}, {Togi}, {Verstraete}, {Witt}, {Wootten}, {Zettergren}, {Zhang}, {Zhang}, \& {Zhen}}]{Chown23}
{Chown}, R., {Sidhu}, A., {Peeters}, E., {et~al.} 2023, arXiv e-prints, arXiv:2308.16733, \dodoi{10.48550/arXiv.2308.16733}

\bibitem[{{Churchwell} {et~al.}(2006){Churchwell}, {Povich}, {Allen}, {Taylor}, {Meade}, {Babler}, {Indebetouw}, {Watson}, {Whitney}, {Wolfire}, {Bania}, {Benjamin}, {Clemens}, {Cohen}, {Cyganowski}, {Jackson}, {Kobulnicky}, {Mathis}, {Mercer}, {Stolovy}, {Uzpen}, {Watson}, \& {Wolff}}]{Churchwell06}
{Churchwell}, E., {Povich}, M.~S., {Allen}, D., {et~al.} 2006, \apj, 649, 759, \dodoi{10.1086/507015}

\bibitem[{{Churchwell} {et~al.}(2009){Churchwell}, {Babler}, {Meade}, {Whitney}, {Benjamin}, {Indebetouw}, {Cyganowski}, {Robitaille}, {Povich}, {Watson}, \& {Bracker}}]{Churchwell09}
{Churchwell}, E., {Babler}, B.~L., {Meade}, M.~R., {et~al.} 2009, \pasp, 121, 213, \dodoi{10.1086/597811}

\bibitem[{{Crocker} {et~al.}(2013){Crocker}, {Calzetti}, {Thilker}, {Aniano}, {Draine}, {Hunt}, {Kennicutt}, {Sandstrom}, \& {Smith}}]{Crocker13}
{Crocker}, A.~F., {Calzetti}, D., {Thilker}, D.~A., {et~al.} 2013, \apj, 762, 79, \dodoi{10.1088/0004-637X/762/2/79}

\bibitem[{{Davis} {et~al.}(2022){Davis}, {Gensior}, {Bureau}, {Cappellari}, {Choi}, {Elford}, {Kruijssen}, {Lelli}, {Liang}, {Liu}, {Ruffa}, {Saito}, {Sarzi}, {Schruba}, \& {Williams}}]{Davis22}
{Davis}, T.~A., {Gensior}, J., {Bureau}, M., {et~al.} 2022, \mnras, 512, 1522, \dodoi{10.1093/mnras/stac600}

\bibitem[{{Deharveng} {et~al.}(2010){Deharveng}, {Schuller}, {Anderson}, {Zavagno}, {Wyrowski}, {Menten}, {Bronfman}, {Testi}, {Walmsley}, \& {Wienen}}]{Deharveng10}
{Deharveng}, L., {Schuller}, F., {Anderson}, L.~D., {et~al.} 2010, \aap, 523, A6, \dodoi{10.1051/0004-6361/201014422}

\bibitem[{{Della Bruna} {et~al.}(2022){Della Bruna}, {Adamo}, {McLeod}, {Smith}, {Savard}, {Robert}, {Sun}, {Amram}, {Bik}, {Blair}, {Long}, {Renaud}, {Walterbos}, \& {Usher}}]{DellaBruna22}
{Della Bruna}, L., {Adamo}, A., {McLeod}, A.~F., {et~al.} 2022, \aap, 666, A29, \dodoi{10.1051/0004-6361/202243395}

\bibitem[{{Draine}(2011)}]{Draine11}
{Draine}, B.~T. 2011, {Physics of the Interstellar and Intergalactic Medium}

\bibitem[{{Draine} \& {Li}(2007)}]{Draine07}
{Draine}, B.~T., \& {Li}, A. 2007, \apj, 657, 810, \dodoi{10.1086/511055}

\bibitem[{{Draine} {et~al.}(2021){Draine}, {Li}, {Hensley}, {Hunt}, {Sandstrom}, \& {Smith}}]{Draine21}
{Draine}, B.~T., {Li}, A., {Hensley}, B.~S., {et~al.} 2021, \apj, 917, 3, \dodoi{10.3847/1538-4357/abff51}

\bibitem[{Draine {et~al.}(2021)Draine, Li, Hensley, Hunt, Sandstrom, \& Smith}]{DraineModels}
Draine, B.~T., Li, A., Hensley, B.~S., {et~al.} 2021, {PAH emission: Dependence on Starlight Spectrum, Intensity, PAH Size Distribution, and PAH Ionization}, V1,  Harvard Dataverse, \dodoi{10.7910/DVN/LPUHIQ}

\bibitem[{{Efremov}(1995)}]{Efremov95}
{Efremov}, Y.~N. 1995, \aj, 110, 2757, \dodoi{10.1086/117728}

\bibitem[{{Egorov} {et~al.}(2023){Egorov}, {Kreckel}, {Sandstrom}, {Leroy}, {Glover}, {Groves}, {Kruijssen}, {Barnes}, {Belfiore}, {Bigiel}, {Blanc}, {Boquien}, {Cao}, {Chastenet}, {Chevance}, {Congiu}, {Dale}, {Emsellem}, {Grasha}, {Klessen}, {Larson}, {Liu}, {Murphy}, {Pan}, {Pessa}, {Pety}, {Rosolowsky}, {Scheuermann}, {Schinnerer}, {Sutter}, {Thilker}, {Watkins}, \& {Williams}}]{Egorov23}
{Egorov}, O.~V., {Kreckel}, K., {Sandstrom}, K.~M., {et~al.} 2023, \apjl, 944, L16, \dodoi{10.3847/2041-8213/acac92}

\bibitem[{{Elmegreen} \& {Elmegreen}(2019)}]{Elmegreen19}
{Elmegreen}, B.~G., \& {Elmegreen}, D.~M. 2019, \apjs, 245, 14, \dodoi{10.3847/1538-4365/ab4903}

\bibitem[{{Elmegreen} {et~al.}(2006){Elmegreen}, {Elmegreen}, {Chandar}, {Whitmore}, \& {Regan}}]{Elmegreen06}
{Elmegreen}, B.~G., {Elmegreen}, D.~M., {Chandar}, R., {Whitmore}, B., \& {Regan}, M. 2006, \apj, 644, 879, \dodoi{10.1086/503797}

\bibitem[{{Emsellem} {et~al.}(2022){Emsellem}, {Schinnerer}, {Santoro}, {Belfiore}, {Pessa}, {McElroy}, {Blanc}, {Congiu}, {Groves}, {Ho}, {Kreckel}, {Razza}, {Sanchez-Blazquez}, {Egorov}, {Faesi}, {Klessen}, {Leroy}, {Meidt}, {Querejeta}, {Rosolowsky}, {Scheuermann}, {Anand}, {Barnes}, {Be{\v{s}}li{\'c}}, {Bigiel}, {Boquien}, {Cao}, {Chevance}, {Dale}, {Eibensteiner}, {Glover}, {Grasha}, {Henshaw}, {Hughes}, {Koch}, {Kruijssen}, {Lee}, {Liu}, {Pan}, {Pety}, {Saito}, {Sandstrom}, {Schruba}, {Sun}, {Thilker}, {Usero}, {Watkins}, \& {Williams}}]{Emsellem22}
{Emsellem}, E., {Schinnerer}, E., {Santoro}, F., {et~al.} 2022, \aap, 659, A191, \dodoi{10.1051/0004-6361/202141727}

\bibitem[{{Engelbracht} {et~al.}(2005){Engelbracht}, {Gordon}, {Rieke}, {Werner}, {Dale}, \& {Latter}}]{Engelbracht05}
{Engelbracht}, C.~W., {Gordon}, K.~D., {Rieke}, G.~H., {et~al.} 2005, \apjl, 628, L29, \dodoi{10.1086/432613}

\bibitem[{{Fahrion} \& {De Marchi}(2023)}]{Fahrion23}
{Fahrion}, K., \& {De Marchi}, G. 2023, \aap, 671, L14, \dodoi{10.1051/0004-6361/202346240}

\bibitem[{{Gaia Collaboration} {et~al.}(2023){Gaia Collaboration}, {Vallenari}, {Brown}, {Prusti}, {de Bruijne}, {Arenou}, {Babusiaux}, {Biermann}, {Creevey}, {Ducourant}, {Evans}, {Eyer}, {Guerra}, {Hutton}, {Jordi}, {Klioner}, {Lammers}, {Lindegren}, {Luri}, {Mignard}, {Panem}, {Pourbaix}, {Randich}, {Sartoretti}, {Soubiran}, {Tanga}, {Walton}, {Bailer-Jones}, {Bastian}, {Drimmel}, {Jansen}, {Katz}, {Lattanzi}, {van Leeuwen}, {Bakker}, {Cacciari}, {Casta{\~n}eda}, {De Angeli}, {Fabricius}, {Fouesneau}, {Fr{\'e}mat}, {Galluccio}, {Guerrier}, {Heiter}, {Masana}, {Messineo}, {Mowlavi}, {Nicolas}, {Nienartowicz}, {Pailler}, {Panuzzo}, {Riclet}, {Roux}, {Seabroke}, {Sordo}, {Th{\'e}venin}, {Gracia-Abril}, {Portell}, {Teyssier}, {Altmann}, {Andrae}, {Audard}, {Bellas-Velidis}, {Benson}, {Berthier}, {Blomme}, {Burgess}, {Busonero}, {Busso}, {C{\'a}novas}, {Carry}, {Cellino}, {Cheek}, {Clementini}, {Damerdji}, {Davidson}, {de Teodoro}, {Nu{\~n}ez Campos}, {Delchambre}, {Dell'Oro}, {Esquej},
  {Fern{\'a}ndez-Hern{\'a}ndez}, {Fraile}, {Garabato}, {Garc{\'\i}a-Lario}, {Gosset}, {Haigron}, {Halbwachs}, {Hambly}, {Harrison}, {Hern{\'a}ndez}, {Hestroffer}, {Hodgkin}, {Holl}, {Jan{\ss}en}, {Jevardat de Fombelle}, {Jordan}, {Krone-Martins}, {Lanzafame}, {L{\"o}ffler}, {Marchal}, {Marrese}, {Moitinho}, {Muinonen}, {Osborne}, {Pancino}, {Pauwels}, {Recio-Blanco}, {Reyl{\'e}}, {Riello}, {Rimoldini}, {Roegiers}, {Rybizki}, {Sarro}, {Siopis}, {Smith}, {Sozzetti}, {Utrilla}, {van Leeuwen}, {Abbas}, {{\'A}brah{\'a}m}, {Abreu Aramburu}, {Aerts}, {Aguado}, {Ajaj}, {Aldea-Montero}, {Altavilla}, {{\'A}lvarez}, {Alves}, {Anders}, {Anderson}, {Anglada Varela}, {Antoja}, {Baines}, {Baker}, {Balaguer-N{\'u}{\~n}ez}, {Balbinot}, {Balog}, {Barache}, {Barbato}, {Barros}, {Barstow}, {Bartolom{\'e}}, {Bassilana}, {Bauchet}, {Becciani}, {Bellazzini}, {Berihuete}, {Bernet}, {Bertone}, {Bianchi}, {Binnenfeld}, {Blanco-Cuaresma}, {Blazere}, {Boch}, {Bombrun}, {Bossini}, {Bouquillon}, {Bragaglia}, {Bramante}, {Breedt},
  {Bressan}, {Brouillet}, {Brugaletta}, {Bucciarelli}, {Burlacu}, {Butkevich}, {Buzzi}, {Caffau}, {Cancelliere}, {Cantat-Gaudin}, {Carballo}, {Carlucci}, {Carnerero}, {Carrasco}, {Casamiquela}, {Castellani}, {Castro-Ginard}, {Chaoul}, {Charlot}, {Chemin}, {Chiaramida}, {Chiavassa}, {Chornay}, {Comoretto}, {Contursi}, {Cooper}, {Cornez}, {Cowell}, {Crifo}, {Cropper}, {Crosta}, {Crowley}, {Dafonte}, {Dapergolas}, {David}, {David}, {de Laverny}, {De Luise}, {De March}, {De Ridder}, {de Souza}, {de Torres}, {del Peloso}, {del Pozo}, {Delbo}, {Delgado}, {Delisle}, {Demouchy}, {Dharmawardena}, {Di Matteo}, {Diakite}, {Diener}, {Distefano}, {Dolding}, {Edvardsson}, {Enke}, {Fabre}, {Fabrizio}, {Faigler}, {Fedorets}, {Fernique}, {Fienga}, {Figueras}, {Fournier}, {Fouron}, {Fragkoudi}, {Gai}, {Garcia-Gutierrez}, {Garcia-Reinaldos}, {Garc{\'\i}a-Torres}, {Garofalo}, {Gavel}, {Gavras}, {Gerlach}, {Geyer}, {Giacobbe}, {Gilmore}, {Girona}, {Giuffrida}, {Gomel}, {Gomez}, {Gonz{\'a}lez-N{\'u}{\~n}ez},
  {Gonz{\'a}lez-Santamar{\'\i}a}, {Gonz{\'a}lez-Vidal}, {Granvik}, {Guillout}, {Guiraud}, {Guti{\'e}rrez-S{\'a}nchez}, {Guy}, {Hatzidimitriou}, {Hauser}, {Haywood}, {Helmer}, {Helmi}, {Sarmiento}, {Hidalgo}, {Hilger}, {H{\l}adczuk}, {Hobbs}, {Holland}, {Huckle}, {Jardine}, {Jasniewicz}, {Jean-Antoine Piccolo}, {Jim{\'e}nez-Arranz}, {Jorissen}, {Juaristi Campillo}, {Julbe}, {Karbevska}, {Kervella}, {Khanna}, {Kontizas}, {Kordopatis}, {Korn}, {K{\'o}sp{\'a}l}, {Kostrzewa-Rutkowska}, {Kruszy{\'n}ska}, {Kun}, {Laizeau}, {Lambert}, {Lanza}, {Lasne}, {Le Campion}, {Lebreton}, {Lebzelter}, {Leccia}, {Leclerc}, {Lecoeur-Taibi}, {Liao}, {Licata}, {Lindstr{\o}m}, {Lister}, {Livanou}, {Lobel}, {Lorca}, {Loup}, {Madrero Pardo}, {Magdaleno Romeo}, {Managau}, {Mann}, {Manteiga}, {Marchant}, {Marconi}, {Marcos}, {Marcos Santos}, {Mar{\'\i}n Pina}, {Marinoni}, {Marocco}, {Marshall}, {Martin Polo}, {Mart{\'\i}n-Fleitas}, {Marton}, {Mary}, {Masip}, {Massari}, {Mastrobuono-Battisti}, {Mazeh}, {McMillan}, {Messina}, {Michalik},
  {Millar}, {Mints}, {Molina}, {Molinaro}, {Moln{\'a}r}, {Monari}, {Mongui{\'o}}, {Montegriffo}, {Montero}, {Mor}, {Mora}, {Morbidelli}, {Morel}, {Morris}, {Muraveva}, {Murphy}, {Musella}, {Nagy}, {Noval}, {Oca{\~n}a}, {Ogden}, {Ordenovic}, {Osinde}, {Pagani}, {Pagano}, {Palaversa}, {Palicio}, {Pallas-Quintela}, {Panahi}, {Payne-Wardenaar}, {Pe{\~n}alosa Esteller}, {Penttil{\"a}}, {Pichon}, {Piersimoni}, {Pineau}, {Plachy}, {Plum}, {Poggio}, {Pr{\v{s}}a}, {Pulone}, {Racero}, {Ragaini}, {Rainer}, {Raiteri}, {Rambaux}, {Ramos}, {Ramos-Lerate}, {Re Fiorentin}, {Regibo}, {Richards}, {Rios Diaz}, {Ripepi}, {Riva}, {Rix}, {Rixon}, {Robichon}, {Robin}, {Robin}, {Roelens}, {Rogues}, {Rohrbasser}, {Romero-G{\'o}mez}, {Rowell}, {Royer}, {Ruz Mieres}, {Rybicki}, {Sadowski}, {S{\'a}ez N{\'u}{\~n}ez}, {Sagrist{\`a} Sell{\'e}s}, {Sahlmann}, {Salguero}, {Samaras}, {Sanchez Gimenez}, {Sanna}, {Santove{\~n}a}, {Sarasso}, {Schultheis}, {Sciacca}, {Segol}, {Segovia}, {S{\'e}gransan}, {Semeux}, {Shahaf}, {Siddiqui}, {Siebert},
  {Siltala}, {Silvelo}, {Slezak}, {Slezak}, {Smart}, {Snaith}, {Solano}, {Solitro}, {Souami}, {Souchay}, {Spagna}, {Spina}, {Spoto}, {Steele}, {Steidelm{\"u}ller}, {Stephenson}, {S{\"u}veges}, {Surdej}, {Szabados}, {Szegedi-Elek}, {Taris}, {Taylor}, {Teixeira}, {Tolomei}, {Tonello}, {Torra}, {Torra}, {Torralba Elipe}, {Trabucchi}, {Tsounis}, {Turon}, {Ulla}, {Unger}, {Vaillant}, {van Dillen}, {van Reeven}, {Vanel}, {Vecchiato}, {Viala}, {Vicente}, {Voutsinas}, {Weiler}, {Wevers}, {Wyrzykowski}, {Yoldas}, {Yvard}, {Zhao}, {Zorec}, {Zucker}, \& {Zwitter}}]{Gaia23}
{Gaia Collaboration}, {Vallenari}, A., {Brown}, A.~G.~A., {et~al.} 2023, \aap, 674, A1, \dodoi{10.1051/0004-6361/202243940}

\bibitem[{{Gordon} {et~al.}(2008){Gordon}, {Engelbracht}, {Rieke}, {Misselt}, {Smith}, \& {Kennicutt}}]{Gordon08}
{Gordon}, K.~D., {Engelbracht}, C.~W., {Rieke}, G.~H., {et~al.} 2008, \apj, 682, 336, \dodoi{10.1086/589567}

\bibitem[{{Gregg} {et~al.}(2024){Gregg}, {Calzetti}, {Adamo}, {Bajaj}, {Ryon}, {Linden}, {Correnti}, {Cignoni}, {Messa}, {Sabbi}, {Gallagher}, {Grasha}, {Pedrini}, {Gutermuth}, {Melinder}, {Kotulla}, {P{\'e}rez}, {Krumholz}, {Bik}, {{\"O}stlin}, {Johnson}, {Bortolini}, {Smith}, {Tosi}, {Maji}, \& {Faustino Vieira}}]{Gregg24}
{Gregg}, B., {Calzetti}, D., {Adamo}, A., {et~al.} 2024, arXiv e-prints, arXiv:2405.09667, \dodoi{10.48550/arXiv.2405.09667}

\bibitem[{{Groves} {et~al.}(2023){Groves}, {Kreckel}, {Santoro}, {Belfiore}, {Zavodnik}, {Congiu}, {Egorov}, {Emsellem}, {Grasha}, {Leroy}, {Scheuermann}, {Schinnerer}, {Watkins}, {Barnes}, {Bigiel}, {Dale}, {Glover}, {Pessa}, {Sanchez-Blazquez}, \& {Williams}}]{Groves23}
{Groves}, B., {Kreckel}, K., {Santoro}, F., {et~al.} 2023, \mnras, 520, 4902, \dodoi{10.1093/mnras/stad114}

\bibitem[{{Guhathakurta} \& {Draine}(1989)}]{Guhathakurta89}
{Guhathakurta}, P., \& {Draine}, B.~T. 1989, \apj, 345, 230, \dodoi{10.1086/167899}

\bibitem[{{Hannon} {et~al.}(2019){Hannon}, {Lee}, {Whitmore}, {Chandar}, {Adamo}, {Mobasher}, {Aloisi}, {Calzetti}, {Cignoni}, {Cook}, {Dale}, {Deger}, {Della Bruna}, {Elmegreen}, {Gouliermis}, {Grasha}, {Grebel}, {Herrero}, {Hunter}, {Johnson}, {Kennicutt}, {Kim}, {Sacchi}, {Smith}, {Thilker}, {Turner}, {Walterbos}, \& {Wofford}}]{Hannon19}
{Hannon}, S., {Lee}, J.~C., {Whitmore}, B.~C., {et~al.} 2019, \mnras, 490, 4648, \dodoi{10.1093/mnras/stz2820}

\bibitem[{{Hannon} {et~al.}(2022){Hannon}, {Lee}, {Whitmore}, {Mobasher}, {Thilker}, {Chandar}, {Adamo}, {Wofford}, {Orozco-Duarte}, {Calzetti}, {Della Bruna}, {Kreckel}, {Groves}, {Barnes}, {Boquien}, {Belfiore}, \& {Linden}}]{Hannon22}
---. 2022, \mnras, 512, 1294, \dodoi{10.1093/mnras/stac550}

\bibitem[{{Hassani} {et~al.}(2023){Hassani}, {Rosolowsky}, {Leroy}, {Boquien}, {Lee}, {Barnes}, {Belfiore}, {Bigiel}, {Cao}, {Chevance}, {Dale}, {Egorov}, {Emsellem}, {Faesi}, {Grasha}, {Kim}, {Klessen}, {Kreckel}, {Kruijssen}, {Larson}, {Meidt}, {Sandstrom}, {Schinnerer}, {Thilker}, {Watkins}, {Whitmore}, \& {Williams}}]{Hassani23}
{Hassani}, H., {Rosolowsky}, E., {Leroy}, A.~K., {et~al.} 2023, \apjl, 944, L21, \dodoi{10.3847/2041-8213/aca8ab}

\bibitem[{{He} {et~al.}(2022){He}, {Wilson}, {Brunetti}, {Finn}, {Bemis}, \& {Johnson}}]{he2022}
{He}, H., {Wilson}, C., {Brunetti}, N., {et~al.} 2022, \apj, 928, 57, \dodoi{10.3847/1538-4357/ac5628}

\bibitem[{{Herrera} {et~al.}(2020){Herrera}, {Pety}, {Hughes}, {Meidt}, {Kreckel}, {Querejeta}, {Saito}, {Lang}, {Jim{\'e}nez-Donaire}, {Pessa}, {Cormier}, {Usero}, {Sliwa}, {Faesi}, {Blanc}, {Bigiel}, {Chevance}, {Dale}, {Grasha}, {Glover}, {Hygate}, {Kruijssen}, {Leroy}, {Rosolowsky}, {Schinnerer}, {Schruba}, {Sun}, \& {Utomo}}]{Herrera20}
{Herrera}, C.~N., {Pety}, J., {Hughes}, A., {et~al.} 2020, \aap, 634, A121, \dodoi{10.1051/0004-6361/201936060}

\bibitem[{{Hollenbach} \& {Tielens}(1999)}]{Hollenbach99}
{Hollenbach}, D.~J., \& {Tielens}, A.~G.~G.~M. 1999, Reviews of Modern Physics, 71, 173, \dodoi{10.1103/RevModPhys.71.173}

\bibitem[{{Hollyhead} {et~al.}(2015){Hollyhead}, {Bastian}, {Adamo}, {Silva-Villa}, {Dale}, {Ryon}, \& {Gazak}}]{Hollyhead15}
{Hollyhead}, K., {Bastian}, N., {Adamo}, A., {et~al.} 2015, \mnras, 449, 1106, \dodoi{10.1093/mnras/stv331}

\bibitem[{{Johnson} {et~al.}(2003){Johnson}, {Indebetouw}, \& {Pisano}}]{Johnson03}
{Johnson}, K.~E., {Indebetouw}, R., \& {Pisano}, D.~J. 2003, \aj, 126, 101, \dodoi{10.1086/375459}

\bibitem[{{Johnson} {et~al.}(2015){Johnson}, {Leroy}, {Indebetouw}, {Brogan}, {Whitmore}, {Hibbard}, {Sheth}, \& {Evans}}]{Johnson15}
{Johnson}, K.~E., {Leroy}, A.~K., {Indebetouw}, R., {et~al.} 2015, \apj, 806, 35, \dodoi{10.1088/0004-637X/806/1/35}

\bibitem[{{Kennicutt} \& {Evans}(2012)}]{Kennicutt12}
{Kennicutt}, R.~C., \& {Evans}, N.~J. 2012, \araa, 50, 531, \dodoi{10.1146/annurev-astro-081811-125610}

\bibitem[{{Krumholz} {et~al.}(2019){Krumholz}, {McKee}, \& {Bland-Hawthorn}}]{Krumholz19}
{Krumholz}, M.~R., {McKee}, C.~F., \& {Bland-Hawthorn}, J. 2019, \araa, 57, 227, \dodoi{10.1146/annurev-astro-091918-104430}

\bibitem[{{Lai} {et~al.}(2020){Lai}, {Smith}, {Baba}, {Spoon}, \& {Imanishi}}]{Lai20}
{Lai}, T. S.~Y., {Smith}, J.~D.~T., {Baba}, S., {Spoon}, H. W.~W., \& {Imanishi}, M. 2020, \apj, 905, 55, \dodoi{10.3847/1538-4357/abc002}

\bibitem[{{Lebouteiller} {et~al.}(2011){Lebouteiller}, {Bernard-Salas}, {Whelan}, {Brandl}, {Galliano}, {Charmandaris}, {Madden}, \& {Kunth}}]{Lebouteiller11}
{Lebouteiller}, V., {Bernard-Salas}, J., {Whelan}, D.~G., {et~al.} 2011, \apj, 728, 45, \dodoi{10.1088/0004-637X/728/1/45}

\bibitem[{{Lee} {et~al.}(2023){Lee}, {Sandstrom}, {Leroy}, {Thilker}, {Schinnerer}, {Rosolowsky}, {Larson}, {Egorov}, {Williams}, {Schmidt}, {Emsellem}, {Anand}, {Barnes}, {Belfiore}, {Be{\v{s}}li{\'c}}, {Bigiel}, {Blanc}, {Bolatto}, {Boquien}, {den Brok}, {Cao}, {Chandar}, {Chastenet}, {Chevance}, {Chiang}, {Congiu}, {Dale}, {Deger}, {Eibensteiner}, {Faesi}, {Glover}, {Grasha}, {Groves}, {Hassani}, {Henny}, {Henshaw}, {Hoyer}, {Hughes}, {Jeffreson}, {Jim{\'e}nez-Donaire}, {Kim}, {Kim}, {Klessen}, {Koch}, {Kreckel}, {Kruijssen}, {Li}, {Liu}, {Lopez}, {Maschmann}, {Chen}, {Meidt}, {Murphy}, {Neumann}, {Neumayer}, {Pan}, {Pessa}, {Pety}, {Querejeta}, {Pinna}, {Rodr{\'\i}guez}, {Saito}, {S{\'a}nchez-Bl{\'a}zquez}, {Santoro}, {Sardone}, {Smith}, {Sormani}, {Scheuermann}, {Stuber}, {Sutter}, {Sun}, {Teng}, {Tre{\ss}}, {Usero}, {Watkins}, {Whitmore}, \& {Razza}}]{Lee23}
{Lee}, J.~C., {Sandstrom}, K.~M., {Leroy}, A.~K., {et~al.} 2023, \apjl, 944, L17, \dodoi{10.3847/2041-8213/acaaae}

\bibitem[{{Leger} {et~al.}(1989{\natexlab{a}}){Leger}, {D'Hendecourt}, {Boissel}, \& {Desert}}]{Leger89b}
{Leger}, A., {D'Hendecourt}, L., {Boissel}, P., \& {Desert}, F.~X. 1989{\natexlab{a}}, \aap, 213, 351

\bibitem[{{Leger} {et~al.}(1989{\natexlab{b}}){Leger}, {D'Hendecourt}, \& {Defourneau}}]{Leger89}
{Leger}, A., {D'Hendecourt}, L., \& {Defourneau}, D. 1989{\natexlab{b}}, \aap, 216, 148

\bibitem[{{Li}(2020)}]{Li20}
{Li}, A. 2020, Nature Astronomy, 4, 339, \dodoi{10.1038/s41550-020-1051-1}

\bibitem[{{Lomaeva} {et~al.}(2022){Lomaeva}, {De Looze}, {Saintonge}, \& {Decleir}}]{Lomaeva22}
{Lomaeva}, M., {De Looze}, I., {Saintonge}, A., \& {Decleir}, M. 2022, \mnras, 517, 3763, \dodoi{10.1093/mnras/stac2940}

\bibitem[{{Mallory} {et~al.}(2022){Mallory}, {Calzetti}, \& {Lin}}]{Mallory22}
{Mallory}, K., {Calzetti}, D., \& {Lin}, Z. 2022, \apj, 933, 156, \dodoi{10.3847/1538-4357/ac7227}

\bibitem[{{Maragkoudakis} {et~al.}(2020){Maragkoudakis}, {Peeters}, \& {Ricca}}]{Maragkoudakis20}
{Maragkoudakis}, A., {Peeters}, E., \& {Ricca}, A. 2020, \mnras, 494, 642, \dodoi{10.1093/mnras/staa681}

\bibitem[{{Martig} {et~al.}(2009){Martig}, {Bournaud}, {Teyssier}, \& {Dekel}}]{Martig09}
{Martig}, M., {Bournaud}, F., {Teyssier}, R., \& {Dekel}, A. 2009, \apj, 707, 250, \dodoi{10.1088/0004-637X/707/1/250}

\bibitem[{{Messa} {et~al.}(2021){Messa}, {Calzetti}, {Adamo}, {Grasha}, {Johnson}, {Sabbi}, {Smith}, {Bajaj}, {Finn}, \& {Lin}}]{Messa21}
{Messa}, M., {Calzetti}, D., {Adamo}, A., {et~al.} 2021, \apj, 909, 121, \dodoi{10.3847/1538-4357/abe0b5}

\bibitem[{{Micelotta} {et~al.}(2010){Micelotta}, {Jones}, \& {Tielens}}]{Micelotta10}
{Micelotta}, E.~R., {Jones}, A.~P., \& {Tielens}, A.~G.~G.~M. 2010, \aap, 510, A36, \dodoi{10.1051/0004-6361/200911682}

\bibitem[{{Osterbrock} \& {Ferland}(2006)}]{Osterbrock06}
{Osterbrock}, D.~E., \& {Ferland}, G.~J. 2006, {Astrophysics of gaseous nebulae and active galactic nuclei}

\bibitem[{{Pathak} {et~al.}(2023){Pathak}, {Leroy}, {Thompson}, {Lopez}, {Belfiore}, {Boquien}, {Dale}, {Glover}, {Klessen}, {Koch}, {Rosolowsky}, {Sandstrom}, {Schinnerer}, {Smith}, {Sun}, {Sutter}, {Williams}, {Bigiel}, {Cao}, {Chastenet}, {Chevance}, {Chown}, {Emsellem}, {Faesi}, {Larson}, {Lee}, {Meidt}, {Ostriker}, {Ramambason}, {Sarbadhicary}, \& {Thilker}}]{Pathak23}
{Pathak}, D., {Leroy}, A.~K., {Thompson}, T.~A., {et~al.} 2023, arXiv e-prints, arXiv:2311.18067, \dodoi{10.48550/arXiv.2311.18067}

\bibitem[{{Peeters} {et~al.}(2023){Peeters}, {Habart}, {Berne}, {Sidhu}, {Chown}, {Van De Putte}, {Trahin}, {Schroetter}, {Canin}, {Alarcon}, {Schefter}, {Khan}, {Pasquini}, {Tielens}, {Wolfire}, {Dartois}, {Goicoechea}, {Maragkoudakis}, {Onaka}, {Pound}, {Vicente}, {Abergel}, {Bergin}, {Bernard-Salas}, {Boersma}, {Bron}, {Cami}, {Cuadrado}, {Dicken}, {Elyajour}, {Fuente}, {Gordon}, {Issa}, {Joblin}, {Kannavou}, {Lacinbala}, {Languignon}, {Le Gal}, {Meshaka}, {Okada}, {Robberto}, {Roellig}, {Schirmer}, {Tabone}, {Zannese}, {Aleman}, {Allamandola}, {Auchettl}, {Baratta}, {Bejaoui}, {Bera}, {Black}, {Boulanger}, {Bouwman}, {Brandl}, {Brechignac}, {Brunken}, {Buragohain}, {Burkhardt}, {Candian}, {Cazaux}, {Cernicharo}, {Chabot}, {Chakraborty}, {Champion}, {Colgan}, {Cooke}, {Coutens}, {Cox}, {Demyk}, {Donovan Meyer}, {Foschino}, {Garcia-Lario}, {Gerin}, {Gottlieb}, {Guillard}, {Gusdorf}, {Hartigan}, {He}, {Herbst}, {Hornekaer}, {Jager}, {Janot-Pacheco}, {Kaufman}, {Kendrew}, {Kirsanova}, {Klaassen}, {Kwok},
  {Labiano}, {Lai}, {Lee}, {Lefloch}, {Le Petit}, {Li}, {Linz}, {Mackie}, {Madden}, {Mascetti}, {McGuire}, {Merino}, {Micelotta}, {Misselt}, {Morse}, {Mulas}, {Neelamkodan}, {Ohsawa}, {Paladini}, {Palumbo}, {Pathak}, {Pendleton}, {Petrignani}, {Pino}, {Puga}, {Rangwala}, {Rapacioli}, {Ricca}, {Roman-Duval}, {Roser}, {Roueff}, {Rouille}, {Salama}, {Sales}, {Sandstrom}, {Sarre}, {Sciamma-O'Brien}, {Sellgren}, {Shenoy}, {Teyssier}, {Thomas}, {Togi}, {Verstraete}, {Witt}, {Wootten}, {Ysard}, {Zettergren}, {Zhang}, {Zhang}, \& {Zhen}}]{Peeters23}
{Peeters}, E., {Habart}, E., {Berne}, O., {et~al.} 2023, arXiv e-prints, arXiv:2310.08720, \dodoi{10.48550/arXiv.2310.08720}

\bibitem[{{Povich} {et~al.}(2007){Povich}, {Stone}, {Churchwell}, {Zweibel}, {Wolfire}, {Babler}, {Indebetouw}, {Meade}, \& {Whitney}}]{Povich07}
{Povich}, M.~S., {Stone}, J.~M., {Churchwell}, E., {et~al.} 2007, \apj, 660, 346, \dodoi{10.1086/513073}

\bibitem[{{Ram{\'\i}rez Alegr{\'\i}a} {et~al.}(2016){Ram{\'\i}rez Alegr{\'\i}a}, {Borissova}, {Chen{\'e}}, {Bonatto}, {Kurtev}, {Amigo}, {Kuhn}, {Gromadzki}, \& {Carballo-Bello}}]{Ramirez16}
{Ram{\'\i}rez Alegr{\'\i}a}, S., {Borissova}, J., {Chen{\'e}}, A.~N., {et~al.} 2016, \aap, 588, A40, \dodoi{10.1051/0004-6361/201526618}

\bibitem[{{Rebolledo} {et~al.}(2015){Rebolledo}, {Wong}, {Xue}, {Leroy}, {Koda}, \& {Donovan Meyer}}]{Rebolledo15}
{Rebolledo}, D., {Wong}, T., {Xue}, R., {et~al.} 2015, \apj, 808, 99, \dodoi{10.1088/0004-637X/808/1/99}

\bibitem[{{Rela{\~n}o} \& {Kennicutt}(2009)}]{Relano09}
{Rela{\~n}o}, M., \& {Kennicutt}, Robert~C., J. 2009, \apj, 699, 1125, \dodoi{10.1088/0004-637X/699/2/1125}

\bibitem[{{Rigby} {et~al.}(2023){Rigby}, {Perrin}, {McElwain}, {Kimble}, {Friedman}, {Lallo}, {Doyon}, {Feinberg}, {Ferruit}, {Glasse}, {Rieke}, {Rieke}, {Wright}, {Willott}, {Colon}, {Milam}, {Neff}, {Stark}, {Valenti}, {Abell}, {Abney}, {Abul-Huda}, {Acton}, {Adams}, {Adler}, {Aguilar}, {Ahmed}, {Albert}, {Alberts}, {Aldridge}, {Allen}, {Altenburg}, {{\'A}lvarez-M{\'a}rquez}, {Alves de Oliveira}, {Andersen}, {Anderson}, {Anderson}, {Argyriou}, {Armstrong}, {Arribas}, {Artigau}, {Arvai}, {Atkinson}, {Bacon}, {Bair}, {Banks}, {Barrientes}, {Barringer}, {Bartosik}, {Bast}, {Baudoz}, {Beatty}, {Bechtold}, {Beck}, {Bergeron}, {Bergkoetter}, {Bhatawdekar}, {Birkmann}, {Blazek}, {Blome}, {Boccaletti}, {B{\"o}ker}, {Boia}, {Bonaventura}, {Bond}, {Bosley}, {Boucarut}, {Bourque}, {Bouwman}, {Bower}, {Bowers}, {Boyer}, {Bradley}, {Brady}, {Braun}, {Breda}, {Bresnahan}, {Bright}, {Britt}, {Bromenschenkel}, {Brooks}, {Brooks}, {Brown}, {Brown}, {Brown}, {Bunker}, {Burger}, {Bushouse}, {Cale}, {Cameron}, {Cameron},
  {Canipe}, {Caplinger}, {Caputo}, {Cara}, {Carey}, {Carniani}, {Carrasquilla}, {Carruthers}, {Case}, {Catherine}, {Chance}, {Chapman}, {Charlot}, {Charlow}, {Chayer}, {Chen}, {Cherinka}, {Chichester}, {Chilton}, {Chonis}, {Clampin}, {Clark}, {Clark}, {Coe}, {Coleman}, {Comber}, {Comeau}, {Connolly}, {Cooper}, {Cooper}, {Coppock}, {Correnti}, {Cossou}, {Coulais}, {Coyle}, {Cracraft}, {Curti}, {Cuturic}, {Davis}, {Davis}, {Dean}, {DeLisa}, {deMeester}, {Dencheva}, {Dencheva}, {DePasquale}, {Deschenes}, {Hunor Detre}, {Diaz}, {Dicken}, {DiFelice}, {Dillman}, {Dixon}, {Doggett}, {Donaldson}, {Douglas}, {DuPrie}, {Dupuis}, {Durning}, {Easmin}, {Eck}, {Edeani}, {Egami}, {Ehrenwinkler}, {Eisenhamer}, {Eisenhower}, {Elie}, {Elliott}, {Elliott}, {Ellis}, {Engesser}, {Espinoza}, {Etienne}, {Etxaluze}, {Falini}, {Feeney}, {Ferry}, {Filippazzo}, {Fincham}, {Fix}, {Flagey}, {Florian}, {Flynn}, {Fontanella}, {Ford}, {Forshay}, {Fox}, {Franz}, {Fu}, {Fullerton}, {Galkin}, {Galyer}, {Garc{\'\i}a Mar{\'\i}n}, {Gardner},
  {Gardner}, {Garland}, {Garrett}, {Gasman}, {Gaspar}, {Gaudreau}, {Gauthier}, {Geers}, {Geithner}, {Gennaro}, {Giardino}, {Girard}, {Giuliano}, {Glassmire}, {Glauser}, {Glazer}, {Godfrey}, {Golimowski}, {Gollnitz}, {Gong}, {Gonzaga}, {Gordon}, {Gordon}, {Goudfrooij}, {Greene}, {Greenhouse}, {Grimaldi}, {Groebner}, {Grundy}, {Guillard}, {Gutman}, {Ha}, {Haderlein}, {Hagedorn}, {Hainline}, {Haley}, {Hami}, {Hamilton}, {Hammel}, {Hansen}, {Harkins}, {Harr}, {Hart}, {Hart}, {Hartig}, {Hashimoto}, {Haskins}, {Hathaway}, {Havey}, {Hayden}, {Hecht}, {Heller-Boyer}, {Henriques}, {Henry}, {Hermann}, {Hernandez}, {Hesman}, {Hicks}, {Hilbert}, {Hines}, {Hoffman}, {Holfeltz}, {Holler}, {Hoppa}, {Hott}, {Howard}, {Howard}, {Hunter}, {Hunter}, {Hurst}, {Husemann}, {Hustak}, {Ilinca Ignat}, {Illingworth}, {Irish}, {Jackson}, {Jahromi}, {Jakobsen}, {James}, {James}, {Januszewski}, {Jenkins}, {Jirdeh}, {Johnson}, {Johnson}, {Jones}, {Jones}, {Jones}, {Jones}, {Jordan}, {Jordan}, {Jurczyk}, {Jurling}, {Kaleida}, {Kalmanson},
  {Kammerer}, {Kang}, {Kao}, {Karakla}, {Kavanagh}, {Kelly}, {Kendrew}, {Kennedy}, {Kenny}, {Keski-kuha}, {Keyes}, {Kidwell}, {Kinzel}, {Kirk}, {Kirkpatrick}, {Kirshenblat}, {Klaassen}, {Knapp}, {Knight}, {Knollenberg}, {Koehler}, {Koekemoer}, {Kovacs}, {Kulp}, {Kumari}, {Kyprianou}, {La Massa}, {Labador}, {Labiano}, {Lagage}, {Lajoie}, {Lallo}, {Lam}, {Lamb}, {Lambros}, {Lampenfield}, {Langston}, {Larson}, {Law}, {Lawrence}, {Lee}, {Leisenring}, {Lepo}, {Leveille}, {Levenson}, {Levine}, {Levy}, {Lewis}, {Lewis}, {Libralato}, {Lightsey}, {Link}, {Liu}, {Lo}, {Lockwood}, {Logue}, {Long}, {Long}, {Loomis}, {Lopez-Caniego}, {Lorenzo Alvarez}, {Love-Pruitt}, {Lucy}, {Luetzgendorf}, {Maghami}, {Maiolino}, {Major}, {Malla}, {Malumuth}, {Manjavacas}, {Mannfolk}, {Marrione}, {Marston}, {Martel}, {Maschmann}, {Masci}, {Masciarelli}, {Maszkiewicz}, {Mather}, {McKenzie}, {McLean}, {McMaster}, {Melbourne}, {Mel{\'e}ndez}, {Menzel}, {Merz}, {Meyett}, {Meza}, {Miskey}, {Misselt}, {Moller}, {Morrison}, {Morse}, {Moseley},
  {Mosier}, {Mountain}, {Mueckay}, {Mueller}, {Mullally}, {Murphy}, {Murray}, {Murray}, {Mustelier}, {Muzerolle}, {Mycroft}, {Myers}, {Myrick}, {Nanavati}, {Nance}, {Nayak}, {Naylor}, {Nelan}, {Nickson}, {Nielson}, {Nieto-Santisteban}, {Nikolov}, {Noriega-Crespo}, {O'Shaughnessy}, {O'Sullivan}, {Ochs}, {Ogle}, {Oleszczuk}, {Olmsted}, {Osborne}, {Ottens}, {Owens}, {Pacifici}, {Pagan}, {Page}, {Park}, {Parrish}, {Patapis}, {Paul}, {Pauly}, {Pavlovsky}, {Pedder}, {Peek}, {Pena-Guerrero}, {Penanen}, {Perez}, {Perna}, {Perriello}, {Phillips}, {Pietraszkiewicz}, {Pinaud}, {Pirzkal}, {Pitman}, {Piwowar}, {Platais}, {Player}, {Plesha}, {Pollizi}, {Polster}, {Pontoppidan}, {Porterfield}, {Proffitt}, {Pueyo}, {Pulliam}, {Quirt}, {Quispe Neira}, {Ramos Alarcon}, {Ramsay}, {Rapp}, {Rapp}, {Rauscher}, {Ravindranath}, {Rawle}, {Regan}, {Reichard}, {Reis}, {Ressler}, {Rest}, {Reynolds}, {Rhue}, {Richon}, {Rickman}, {Ridgaway}, {Ritchie}, {Rix}, {Robberto}, {Robinson}, {Robinson}, {Robinson}, {Rock}, {Rodriguez}, {Rodriguez
  Del Pino}, {Roellig}, {Rohrbach}, {Roman}, {Romelfanger}, {Rose}, {Roteliuk}, {Roth}, {Rothwell}, {Rowlands}, {Roy}, {Royer}, {Royle}, {Rui}, {Rumler}, {Runnels}, {Russ}, {Rustamkulov}, {Ryden}, {Ryer}, {Sabata}, {Sabatke}, {Sabbi}, {Samuelson}, {Sapp}, {Sappington}, {Sargent}, {Sauer}, {Scheithauer}, {Schlawin}, {Schlitz}, {Schmitz}, {Schneider}, {Schreiber}, {Schulze}, {Schwab}, {Scott}, {Sembach}, {Shanahan}, {Shaughnessy}, {Shaw}, {Shawger}, {Shay}, {Sheehan}, {Shen}, {Sherman}, {Shiao}, {Shih}, {Shivaei}, {Sienkiewicz}, {Sing}, {Sirianni}, {Sivaramakrishnan}, {Skipper}, {Sloan}, {Slocum}, {Slowinski}, {Smith}, {Smith}, {Smith}, {Smith}, {Snyder}, {Soh}, {Sohn}, {Soto}, {Spencer}, {Stallcup}, {Stansberry}, {Starr}, {Starr}, {Stewart}, {Stiavelli}, {Straughn}, {Strickland}, {Stys}, {Summers}, {Sun}, {Sunnquist}, {Swade}, {Swam}, {Swaters}, {Swoish}, {Taylor}, {Taylor}, {Te Plate}, {Tea}, {Teague}, {Telfer}, {Temim}, {Thatte}, {Thompson}, {Thompson}, {Thomson}, {Tikkanen}, {Tippet}, {Todd}, {Toolan},
  {Tran}, {Trejo}, {Truong}, {Tsukamoto}, {Tustain}, {Tyra}, {Ubeda}, {Underwood}, {Uzzo}, {Van Campen}, {Vandal}, {Vandenbussche}, {Vila}, {Volk}, {Wahlgren}, {Waldman}, {Walker}, {Wander}, {Warfield}, {Warner}, {Wasiak}, {Watkins}, {Weaver}, {Weilert}, {Weiser}, {Weiss}, {Weissman}, {Welty}, {West}, {Wheate}, {Wheatley}, {Wheeler}, {White}, {Whiteaker}, {Whitehouse}, {Whiteleather}, {Whitman}, {Williams}, {Willmer}, {Willoughby}, {Wilson}, {Wirth}, {Wislowski}, {Wolf}, {Wolfe}, {Wolff}, {Workman}, {Wright}, {Wu}, {Wu}, {Wymer}, {Yates}, {Yeager}, {Yeates}, {Yerger}, {Yoon}, {Young}, {Yu}, {Zak}, {Zeidler}, {Zhou}, {Zielinski}, {Zincke}, \& {Zonak}}]{Rigby23}
{Rigby}, J., {Perrin}, M., {McElwain}, M., {et~al.} 2023, \pasp, 135, 048001, \dodoi{10.1088/1538-3873/acb293}

\bibitem[{{Sandstrom} {et~al.}(2012){Sandstrom}, {Bolatto}, {Bot}, {Draine}, {Ingalls}, {Israel}, {Jackson}, {Leroy}, {Li}, {Rubio}, {Simon}, {Smith}, {Stanimirovi{\'c}}, {Tielens}, \& {van Loon}}]{Sandstrom12}
{Sandstrom}, K.~M., {Bolatto}, A.~D., {Bot}, C., {et~al.} 2012, \apj, 744, 20, \dodoi{10.1088/0004-637X/744/1/20}

\bibitem[{{Sandstrom} {et~al.}(2023){Sandstrom}, {Koch}, {Leroy}, {Rosolowsky}, {Emsellem}, {Smith}, {Egorov}, {Williams}, {Larson}, {Lee}, {Schinnerer}, {Thilker}, {Barnes}, {Belfiore}, {Bigiel}, {Blanc}, {Bolatto}, {Boquien}, {Cao}, {Chastenet}, {Chevance}, {Chiang}, {Dale}, {Faesi}, {Glover}, {Grasha}, {Groves}, {Hassani}, {Henshaw}, {Hughes}, {Kim}, {Klessen}, {Kreckel}, {Kruijssen}, {Lopez}, {Liu}, {Meidt}, {Murphy}, {Pan}, {Querejeta}, {Saito}, {Sardone}, {Sormani}, {Sutter}, {Usero}, \& {Watkins}}]{Sandstrom23b}
{Sandstrom}, K.~M., {Koch}, E.~W., {Leroy}, A.~K., {et~al.} 2023, \apjl, 944, L8, \dodoi{10.3847/2041-8213/aca972}

\bibitem[{{Schroetter} {et~al.}(2024){Schroetter}, {Bern{\'e}}, {Joblin}, {Canin}, {Chown}, {Sidhu}, {Habart}, {Peeters}, {Lai}, {Candian}, {Chakraborty}, {Petrignani}, {Trahin}, {Van De Putte}, \& {Alarc{\'o}n}}]{Schroetter24}
{Schroetter}, I., {Bern{\'e}}, O., {Joblin}, C., {et~al.} 2024, \aap, 685, A78, \dodoi{10.1051/0004-6361/202348974}

\bibitem[{{Seok} {et~al.}(2014){Seok}, {Hirashita}, \& {Asano}}]{Seok14}
{Seok}, J.~Y., {Hirashita}, H., \& {Asano}, R.~S. 2014, \mnras, 439, 2186, \dodoi{10.1093/mnras/stu120}

\bibitem[{{Smith} {et~al.}(2007){Smith}, {Draine}, {Dale}, {Moustakas}, {Kennicutt}, {Helou}, {Armus}, {Roussel}, {Sheth}, {Bendo}, {Buckalew}, {Calzetti}, {Engelbracht}, {Gordon}, {Hollenbach}, {Li}, {Malhotra}, {Murphy}, \& {Walter}}]{Smith07}
{Smith}, J.~D.~T., {Draine}, B.~T., {Dale}, D.~A., {et~al.} 2007, \apj, 656, 770, \dodoi{10.1086/510549}

\bibitem[{{Stutz}(2018)}]{Stutz18}
{Stutz}, A.~M. 2018, \mnras, 473, 4890, \dodoi{10.1093/mnras/stx2629}

\bibitem[{{Sutter} {et~al.}(2024){Sutter}, {Sandstrom}, {Chastenet}, {Leroy}, {Koch}, {Williams}, {Chown}, {Belfiore}, {Bigiel}, {Boquien}, {Cao}, {Chevance}, {Dale}, {Egorov}, {Glover}, {Groves}, {Klessen}, {Kreckel}, {Larson}, {Oakes}, {Pathak}, {Ramambason}, {Rosolowsky}, \& {Watkins}}]{Sutter24}
{Sutter}, J., {Sandstrom}, K., {Chastenet}, J., {et~al.} 2024, arXiv e-prints, arXiv:2405.15102, \dodoi{10.48550/arXiv.2405.15102}

\bibitem[{{Tielens}(2008)}]{Tielens08}
{Tielens}, A.~G.~G.~M. 2008, \araa, 46, 289, \dodoi{10.1146/annurev.astro.46.060407.145211}

\bibitem[{{Tully} {et~al.}(2009){Tully}, {Rizzi}, {Shaya}, {Courtois}, {Makarov}, \& {Jacobs}}]{Tully09}
{Tully}, R.~B., {Rizzi}, L., {Shaya}, E.~J., {et~al.} 2009, \aj, 138, 323, \dodoi{10.1088/0004-6256/138/2/323}

\bibitem[{{Verstraete} {et~al.}(2001){Verstraete}, {Pech}, {Moutou}, {Sellgren}, {Wright}, {Giard}, {L{\'e}ger}, {Timmermann}, \& {Drapatz}}]{Verstraete01}
{Verstraete}, L., {Pech}, C., {Moutou}, C., {et~al.} 2001, \aap, 372, 981, \dodoi{10.1051/0004-6361:20010515}

\bibitem[{{Watkins} {et~al.}(2023){Watkins}, {Barnes}, {Henny}, {Kim}, {Kreckel}, {Meidt}, {Klessen}, {Glover}, {Williams}, {Keller}, {Leroy}, {Rosolowsky}, {Lee}, {Anand}, {Belfiore}, {Bigiel}, {Blanc}, {Boquien}, {Cao}, {Chandar}, {Chen}, {Chevance}, {Congiu}, {Dale}, {Deger}, {Egorov}, {Emsellem}, {Faesi}, {Grasha}, {Groves}, {Hassani}, {Henshaw}, {Herrera}, {Hughes}, {Jeffreson}, {Jim{\'e}nez-Donaire}, {Koch}, {Kruijssen}, {Larson}, {Liu}, {Lopez}, {Pessa}, {Pety}, {Querejeta}, {Saito}, {Sandstrom}, {Scheuermann}, {Schinnerer}, {Sormani}, {Stuber}, {Thilker}, {Usero}, \& {Whitmore}}]{Watkins23}
{Watkins}, E.~J., {Barnes}, A.~T., {Henny}, K., {et~al.} 2023, \apjl, 944, L24, \dodoi{10.3847/2041-8213/aca6e4}

\bibitem[{{Watson} {et~al.}(2008){Watson}, {Povich}, {Churchwell}, {Babler}, {Chunev}, {Hoare}, {Indebetouw}, {Meade}, {Robitaille}, \& {Whitney}}]{Watson08}
{Watson}, C., {Povich}, M.~S., {Churchwell}, E.~B., {et~al.} 2008, \apj, 681, 1341, \dodoi{10.1086/588005}

\bibitem[{{Whitmore} {et~al.}(2011){Whitmore}, {Chandar}, {Kim}, {Kaleida}, {Mutchler}, {Stankiewicz}, {Calzetti}, {Saha}, {O'Connell}, {Balick}, {Bond}, {Carollo}, {Disney}, {Dopita}, {Frogel}, {Hall}, {Holtzman}, {Kimble}, {McCarthy}, {Paresce}, {Silk}, {Trauger}, {Walker}, {Windhorst}, \& {Young}}]{Whitmore11}
{Whitmore}, B.~C., {Chandar}, R., {Kim}, H., {et~al.} 2011, \apj, 729, 78, \dodoi{10.1088/0004-637X/729/2/78}

\bibitem[{{Xie} \& {Ho}(2019)}]{Xie19}
{Xie}, Y., \& {Ho}, L.~C. 2019, \apj, 884, 136, \dodoi{10.3847/1538-4357/ab4200}

\bibitem[{{Xie} \& {Ho}(2022)}]{Xie22}
---. 2022, \apj, 925, 218, \dodoi{10.3847/1538-4357/ac32e2}

\end{thebibliography}
\bibliographystyle{aasjournal}



\end{document}